\documentclass[prd,twocolumn,preprintnumbers,floatfix]{revtex4-1}

\usepackage{amsthm}
\usepackage{mathtools}
\usepackage{physics}
\usepackage{xcolor}
\usepackage{graphicx}
\usepackage[left=23mm,right=13mm,top=35mm,columnsep=15pt]{geometry} 
\usepackage{adjustbox}
\usepackage{placeins}
\usepackage[T1]{fontenc}
\usepackage{lipsum}
\usepackage[utf8]{inputenc}
\usepackage{csquotes}
\usepackage[english]{babel}
\usepackage{amsmath}
\usepackage{titlesec}
\usepackage{soul}
\usepackage{soulutf8}
\usepackage{comment}
\sethlcolor{white}
\titlespacing{\section}{0pt}{*5}{*2}
\titlespacing{\subsection}{0pt}{*3}{*2}
\usepackage{enumitem}
\usepackage{hyperref}
\hypersetup{pdftitle={Article}, pdfauthor={Author}}

\usepackage{array,multirow,xcolor,colortbl}
\usepackage{adjustbox}
\definecolor{amber}{HTML}{FAD7A0}
\definecolor{redDark}{HTML}{EA9999} 
\definecolor{greenMed}{HTML}{B7DDBF}
\newcolumntype{C}[1]{>{\centering\arraybackslash}m{#1}}

\begin{document}

\bibliographystyle{apsrev4-1}

\title{MaxWave Signal: Rapid, coherent maximum likelihood \\ wavelet reconstruction of transient signals in gravitational wave data}
\author{Sudhi Mathur}
    \affiliation{eXtreme Gravity Institute, Department of Physics, Montana State University, Bozeman, Montana 59717 USA}

\author{Neil J. Cornish}
    \affiliation{eXtreme Gravity Institute, Department of Physics, Montana State University, Bozeman, Montana 59717 USA}
    
\date{June 22, 2026}

\begin{abstract}
\noindent Advances in gravitational-wave detector sensitivity have increased the rate of transient signal detections, demanding faster automated analysis. We extend MaxWave, a fast maximum likelihood wavelet reconstruction algorithm, to perform coherent multi-detector signal reconstruction and glitch rejection. We coherently search for a common set of wavelets modeling the signal in all detectors. Multi-detector data are aligned using z-statistic time and phase offsets and amplitude scalings relative to the dominant reconstruction, as well as adaptive noise weightings derived from a geometrically averaged noise spectrum. By aligning and weighting individual detectors, we form a synthetic detector that amplifies non-Gaussian features, down-weights noisy detectors, and preserves Gaussian noise statistics. We extract the coherent signal using this synthetic detector, improving sensitivity to weak events while rejecting coincident glitches that lack consistent phase and amplitude evolution. Our algorithm provides real-time, low-latency, model-independent signal reconstructions, safely denoises gravitational wave data without removing transient signals, and can complement existing burst search and reconstruction frameworks through a fundamentally distinct approach, strengthening detection confidence and improving sensitivity to diverse signal morphologies.

\end{abstract}


\maketitle

\section{INTRODUCTION} \label{sec:Introduction}

Advances \cite{LVK, AdLIGO-O4-2025, Soni_2025, Di_Pace_2021} in gravitational-wave detector \cite{LVK, AdLIGO-2015, Abbott_2009, Acernese_2008} sensitivity have increased the rate of transient signal detections, demanding faster automated reconstructions that are robust against non-Gaussian noise artifacts called "glitches". Although well-modeled transients such as compact binary coalescences (CBCs) with low mass ratios and spins are optimally detected and characterized using matched-filtered searches, a broad range of sources, including extreme-mass-ratio or highly precessing binaries, core-collapse supernovae, post-merger remnants, and other unexpected astrophysical phenomena, can produce signals with unknown morphology. For such signals, model-independent multi-detector coherent burst analyses are crucial. 

We extend MaxWave \cite{MaxWave_Glitch}, a fast maximum likelihood model-independent wavelet reconstruction algorithm, to perform coherent multi-detector transient signal reconstruction and glitch rejection. Our model, MaxWave signal, is designed to complement existing coherent burst search frameworks such as X-Pipeline \cite{Xpipeline} and Coherent Waveburst (cWB) \cite{cwb1, cwb2}, which has been developed over the last 20 years. Burst searches that employ distinct time-frequency representations, likelihood functions, and reconstruction bases extract complementary signal information and exhibit varying sensitivities to glitch morphologies. Thus, a diverse ecosystem of burst search pipelines is essential to strengthen detection confidence and flag discrepancies indicative of data quality issues or source complexity. While cWB constructs coherence through detector antenna pattern projections, phase-transformed polarization response, and ellipticity-polarization constraints \cite{cwb1, cwb2}, X-Pipeline through coherent and incoherent energy channels projected in the dominant polarization frame \cite{Xpipeline}, and DeepExtractor \cite{DeepExtractor} and AWaRe \cite{AWaRe1, AWaRe2} through learned single-detector noise and signal reconstructions respectively, we adopt a conceptually distinct reconstruction-based approach.

MaxWave signal uses z-statistic time and phase offsets and amplitude scalings \cite{FINDCHIRP, Neil_QuickCBC}, computed against the highest signal-to-noise ratio (SNR) reconstruction, as well as adaptive noise weightings derived from a geometrically averaged noise spectrum, to weight and align multi-detector data. This creates a single coherent synthetic detector that amplifies non-Gaussian features, down-weights noisy detectors, and preserves Gaussian noise statistics. We reconstruct the coherent signal using this synthetic detector, improving sensitivity to weak events while rejecting coincident glitches that lack consistent phase and amplitude evolution. Using a combined detector stream approach and approximate likelihood maximization on a coherent detector, MaxWave signal implicitly encodes network diagnostics and provides a novel framework for low-latency signal and glitch analysis. Although beyond the scope of this work, our model can be extended  to a complete, stand-alone detection pipeline by further quantifying the detection statistics and comparing it against the performance of existing burst search algorithms.

Through the coherent multi-detector reconstruction, MaxWave signal can accelerate the convergence time for the \textit{BayesWave} signal model \cite{Cornish_2021} by starting the sampler's reversible jump Markov Chain Monte Carlo (RJMCMC) chains near a good initial solution. Currently, the \textit{BayesWave} signal model's chains are initialized by crudely drawing the wavelet intrinsic and extrinsic parameters from broad flat prior distributions. While the \textit{BayesWave} noise model \cite{Cornish_2015} benefits from an approximate maximum likelihood initialization \cite{Cornish_2021, MaxWave_Glitch}, this solution is not suitable for initializing the signal model. The noise model's glitch wavelets are constructed independently in each detector's reference frame, whereas the signal model builds a single geocentric waveform projected coherently onto the detector network. It employs a fundamentally different parameterization that requires additional extrinsic parameters, including sky location, polarization angle, and ellipticity. MaxWave signal's coherent multi-detector reconstruction directly provides this geocentric starting point, and seeds the RJMCMC chains near a good initial solution in the hard to sample, coupled, multimodal parameter space where sky location and wavelet time shifts are jointly constrained across detectors. Our model can also safely denoise gravitational wave data without removing transient signals.  

In the following sections, we present the formulation, implementation, and performance of the MaxWave signal model. Sec. \ref{sec:Methods} describes our methodology. We construct a coherent synthetic detector and outline glitch rejection tests. Sec. \ref{sec:Results} presents our results. We demonstrate improvements in multi-detector signal reconstructions, residuals, and injected signal recovery using the coherent detector. We compare MaxWave signal to \textit{BayesWave} \cite{Cornish_2021} in terms of reconstruction accuracy across various SNRs and binary mass ratios. We present glitch rejection statistics, calculate operations cost and runtime, and analyze 14 hours of real LIGO data for a two-detector network. We also apply our model to two binary black hole events GW191109\_010717 and GW200129\_065458 (subsequently referred to as GW191109 and GW200129) from the third observing run (O3) \cite{GWTC-3, GWOSC_O1_O3} where the parameter estimation of the gravitational wave signals is affected by glitches \cite{Udall_2024, Payne_2022, Davis_2022}. Sec. \ref{sec:Conclusion} summarizes our conclusions. Sec. \ref{sec:Future Directions and Scope} discusses broader applications such as extending our model to a stand-alone low-latency burst search pipeline. 

\section{METHODS} \label{sec:Methods}

\begin{figure}[b]
    \centering
    \includegraphics[width=\linewidth]{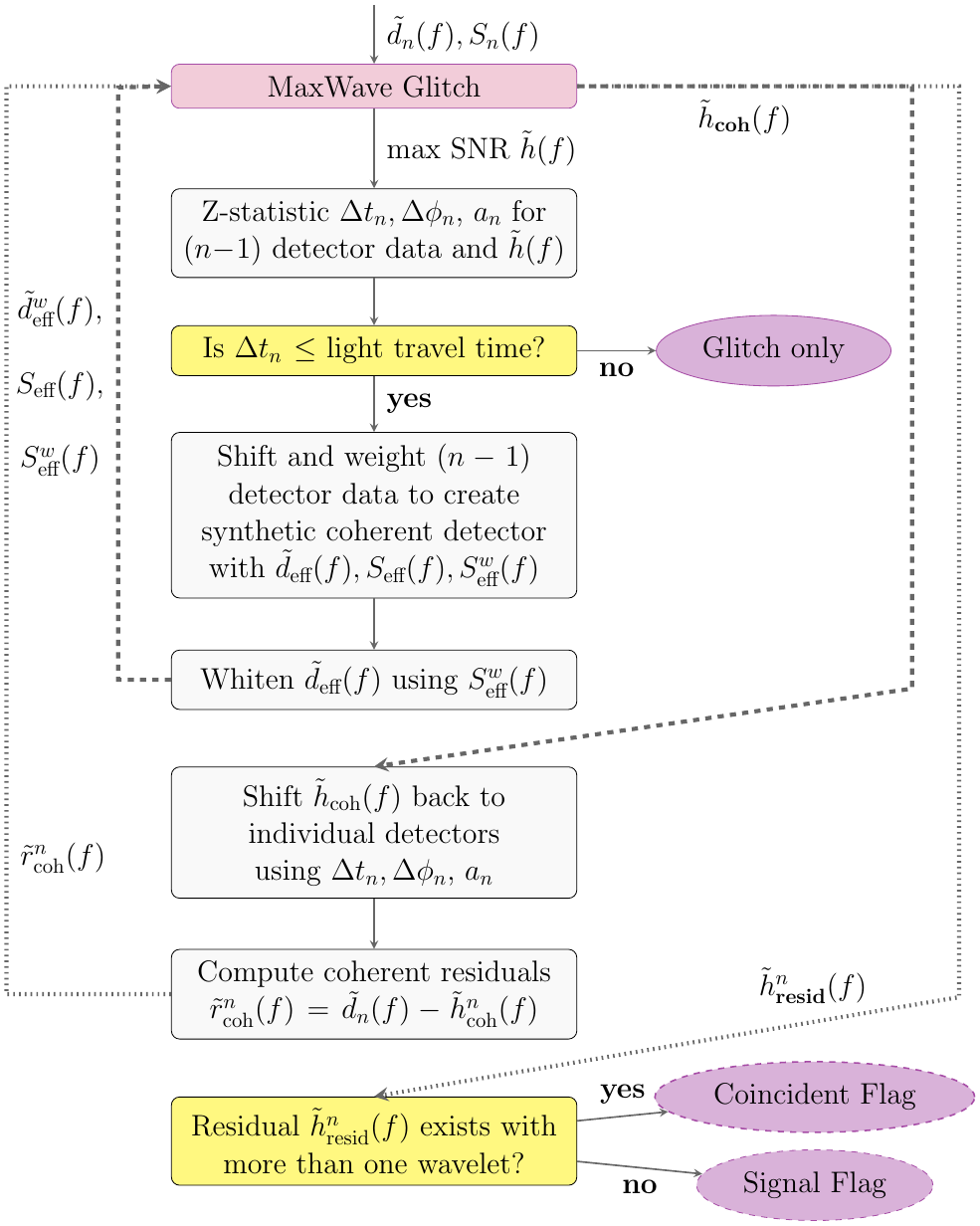}
    \caption{Model for the coherent multi-detector MaxWave signal reconstruction. MaxWave glitch model is run independently on all $n$ detectors and highest-SNR reconstruction $\tilde{h}(f)$ is selected as a reference. Z-statistic time and phase shifts, and amplitude scalings $(\Delta t_n, \Delta \phi_n, a_n)$ are computed for the remaining $(n-1)$ detectors and if $\Delta t_n$ is consistent with the inter-detector light-travel time, the individual detectors are coherently aligned, scaled and combined into a synthetic detector. MaxWave is re-run on this synthetic detector to obtain a coherent reconstruction $\tilde{h}_\text{coh}(f)$. The coherent reconstruction is shifted and scaled back to individual detectors $\tilde{h}_\text{coh}^n(f)$ and used to compute residuals $\tilde{r}_\text{coh}^n(f) = \tilde{d}_n(f) - \tilde{h}_\text{coh}^n(f)$. MaxWave is re-run on these residuals, and any reconstruction $\tilde{h}_\text{resid}^n(f)$ with a significant number of wavelets is indicative of incoherent residuals and receives a "coincident event non-removal" flag instead of a "signal non-removal" flag.}
    \label{fig:Signal_flowchart}
\end{figure}

For coherent multi-detector reconstruction, our algorithm (Fig. \ref{fig:Signal_flowchart}) searches for a common wavelet representation of the signal across the detector network. After running the MaxWave glitch model \cite{MaxWave_Glitch} independently on data from all $n$ detectors and rendering individual detector reconstructions, we identify the detector $n_\text{max}$ with the highest SNR reconstruction $\tilde{h}(f)$. Using $\tilde{h}(f)$ as a reference, we compute z-statistic \cite{FINDCHIRP, Neil_QuickCBC} time and phase shifts ($\Delta t_n$ and $\Delta \phi_n$), and amplitude scalings $a_n$ with respect to the detector strain from the other $(n-1)$ detectors. The z-statistic defines the complex cross-correlation between $\tilde{h}(f)$ and $\tilde{d}_n(f)$ as a function of time delay $\Delta t$,
\begin{equation}
    z_n(\Delta t) = 4\int_0^\infty \frac{\tilde{d}_n(f)\,\tilde{h}^*(f)}{S_n(f)}\,e^{2\pi i f \Delta t}
\end{equation}
where the peaks of $|z_n(\Delta t)|$, the phases $\arg[z_n(\Delta t_n)]$, and the ratios of $|z_n(\Delta t_n)|$ to the wavelet inner product $(h|h)^{1/2}$ give us the z-statistic time shifts $\Delta t_n$, phase shifts $\Delta\phi_n$, and amplitude scalings $a_n$, respectively \cite{FINDCHIRP}. If the z-statistic time shift between $\tilde{h}(f)$ and any of the $(n-1)$ detector strains is less than the inter-detector light travel time, and at least one of the $(n-1)$ detectors within the light travel time has SNR $\geq$ 5, we proceed to coherently align them. This approach allows us to dig deeper and extract additional signal information even from detectors with weak signals and no initial reconstruction. 

We weight and align individual detectors that pass the z-statistic light travel time test to construct a coherent synthetic detector that amplifies non-Gaussian features and preserves Gaussian noise statistics (further described in Sec. \ref{subsec:comb_multi}). We re-run MaxWave glitch model on this synthetic detector and produce a coherent signal reconstruction $\tilde{h}_\text{coh}(f)$ that captures the common set of wavelets across our multi-detector network.

Unlike signals, false alarms such as coincident glitches are characterized by independent non-Gaussian structure in each detector and produce incoherent multi-detector reconstructions. Even if glitches occur within the light travel time, their distinct time–frequency morphology is reflected in the synthetic detector response and in the resulting reconstruction $\tilde{h}_\text{coh}(f)$. To catch false alarms and cases such as coincident glitches, or glitch-signal overlaps that might produce z-statistic time-shifts within the inter-detector light travel time, we shift and scale the coherent reconstruction back to the individual detectors $\tilde{h}_\text{coh}^n(f)$ and compute the coherent reconstruction residuals $\tilde{r}_\text{coh}^n(f) = \tilde{d}_n(f) - \tilde{h}_\text{coh}^n(f)$. We re-run the MaxWave glitch model on these residuals to capture any significant incoherent power. For signals, the reconstruction $\tilde{h}_\text{resid}(f)$ extracted from the coherent residuals should either not exist or have occasional low-SNR single wavelets for cases where the initial reconstruction was noisy and the coherent reconstruction removed the noise-induced wavelet. Thus, any reconstruction $\tilde{h}_\text{resid}(f)$ with more than one wavelet is flagged as a "coincident event non-removal" instead of a "signal non-removal", where "non-removal" signifies that the event passed the z-statistic light travel time test and should not be removed when denoising the gravitational wave data. We note that the classification of non-removals as coincident events or signals based on the coherent residuals is not conclusive, but provisional (indicated by the dotted border in Fig. \ref{fig:Signal_flowchart}).

\subsection{Constructing the synthetic coherent detector}\label{subsec:comb_multi}

We construct a synthetic coherent detector by time-shifting, phase-rotating, and amplitude-scaling the strain data from individual detectors if an event passes the z-statistic light travel time consistency test, and illustrate its statistical properties, including preserved Gaussian noise background and amplified coherent non-Gaussian features.

\subsubsection{Two Detector Case}
The standard inner product \cite{Finn_1992} for $K$ detectors has the form
\begin{equation}
    (d|h) = \sum_{k=1}^K 2 \int df \left[ \frac{d_k h_k^* + d_k^* h_k}{S_k} \right]
\end{equation}
where the frequency dependence of the detector strain data $d_k(f)$, waveform reconstruction templates $h_k(f)$ and power spectral densities $S_k(f)$ is omitted for clarity. 

In a standard template based analysis, the projection of the geocentric reference template $h$ to each individual detector $h_k$ depends on the sky location of the source, detector antenna patterns, and the two general relativistic gravitational wave polarization modes $h_+(f)$ and $h_\times(f)$ \cite{Cornish_2021}.

Currently, we restrict the MaxWave signal model to elliptical polarizations
\begin{equation}
\begin{aligned}
    h_+ & = \sum_n\Psi(f;t_0^n, f_0^n, \tau^n, \mathcal{A}^n,\phi_0^n) \\
    h_\times & = i\epsilon h_+
\end{aligned}
\end{equation}
where $\mathcal{A}^n$ are the wavelet amplitudes, $t_0^n$ the central times, $f_0^n$ the central frequencies,  $\phi_0^n$ the phases at $t = t_0^n$, $\tau^n$ the time extent related to the quality factors as $\tau^n = Q^n/2\pi f_0^n$, and $\epsilon$ is the ellipticity ranging from $\epsilon=0$ linearly polarized to $\epsilon=1$ circularly polarized signals. 

This elliptical polarization assumption does not hold for CBCs with spin-precession or strong higher-order modes as their polarization content changes with time. For highly precessing CBC signals, relaxing the elliptical polarization assumption to generic polarizations modestly improves the 90\% credible signal reconstruction intervals (Fig. 1 in \cite{Cornish_2021}). Generic gravitational wave bursts are also not expected to have any special polarization relation between the two modes, and the polarization restriction is suboptimal for detecting unpolarized signals such as white noise bursts \cite{Cornish_2021}. However, the assumption still works well for our low-latency model. It allows us to safely denoise GW data, and provide a good initial, coherent, multi-detector solution that speeds up convergence of computationally intensive algorithms such as the \textit{BayesWave} signal model, which can then extend to generic polarizations.

We estimate the $h_k$ by running the MaxWave glitch model on individual detectors. Consider a two-detector network and suppose that $h_1$ is the higher SNR reconstruction. We use z-statistic \cite{FINDCHIRP, Neil_QuickCBC} time and phase differences between $h_1$ and the $d_2$ detector data to shift the data in the lower SNR detector in time and in overall phase. Assuming elliptical polarization, after this alignment, the $h$'s only differ by an overall amplitude scaling, such that $h = h_1 = \alpha h_2$ and
 
\begin{equation}
    (d|h) = 2 \int df \left[ \frac{d_1 h^* + d_1^* h}{S_1} + \frac{1}{\alpha} \frac{d'_2 h^* + {d'}_2^* h}{S_2} \right]
\end{equation}
where $d'_2(f)$ is the time and phase shifted data in lower SNR detector. We can re-write this as
\begin{equation}
    (d|h) = 2 \int df \left[ \frac{d_{\text{eff}} h^* + d_{\text{eff}}^* h}{S_{\text{eff}}} \right]
    \label{Eq:deff inner product}
\end{equation}
where the detector strain of the synthetic coherent detector $d_\text{eff}(f)$ is given as
\begin{equation}
    d_{\text{eff}} = \sqrt{\frac{S_2}{S_1}} d_1 + \frac{1}{\alpha} \sqrt{\frac{S_1}{S_2}} d'_2
\end{equation}
and the effective power spectral density $S_\text{eff}(f)$ used to weight the standard inner product is
\begin{equation}
    S_{\text{eff}} = \sqrt{S_1 S_2}
\end{equation}
The relative amplitude scaling inferred from z-statistic for the high SNR template $h_1$ and lower SNR detector data $d_2$ impacts the rescaling of data in the shifted detector $d'_2$. In the limit that one detector is much more sensitive than the other, the inner product in Eq. \ref{Eq:deff inner product} reduces to that for the single detector case. The generalization to $K$ detectors is direct.

Note that $S_{\text{eff}}(f)$ does not whiten the data, but is used solely for calculating inner products. To whiten the data, we compute $S^w_{\text{eff}} = \text{E}[(d_{\text{eff}})^2]$ with no cross terms to get
\begin{equation}
    S^w_{\text{eff}} = \frac{S_2}{S_1} \text{E}[(d_1)^2] + \frac{1}{\alpha^2} \frac{S_1}{S_2} \text{E}[(d'_2)^2]
\end{equation}
Thus the power spectral density that whitens the effective data is given by 
\begin{equation}
     S^w_{\text{eff}} =  S_2 + \frac{1}{\alpha^2} S_1
\end{equation}
which is a Gaussian random variable. Thus, whitening with $S^w_{\text{eff}}(f)$ preserves Gaussian noise statistics in the effective whitened data $d^w_\text{eff}(f)$ of the synthetic coherent detector.

\subsubsection{Generalizing for multiple detectors}

Let $h= a_1 h_1$ be the highest SNR reconstruction among $n$ single detector reconstructions $h_n$ such that $a_1 \approx 1$. We use the time and phase shifts between $h$ and $d_n$ and to align the data in all subsequent detectors to the detector with highest SNR reconstruction. After this, assuming elliptical polarization, $h_n$ only differ by an overall amplitude scaling such that 
\begin{equation}
    h = a_1 h_1 = a_2 h_2 = ... = a_n h_n
\end{equation}
Substituting $h_n = h/a_n$ in the standard inner product equation
\begin{equation}
    (d|h) = 2 \int df \sum_n \frac{d_n h_n^* + {d}_n^* h_n}{S_n}
\end{equation}
we get
\begin{equation}
    (d|h) = 2 \int df \sum_n \frac{1}{a_n} \frac{d'_n h^* + {d'}_n^* h}{S_n}
    \label{Eq:d_n inner product sum}
\end{equation}
where $d'_n(f)$ is the time and phase shifted data in detector $n$. Expanding this further
\begin{align}
    (d|h) = 2 \int df \Bigg[ \frac{1}{a_1} \frac{d_1 h^* + d_1^* h}{S_1}  & + \frac{1}{a_2} \frac{d'_2 h^* + d_2^* h}{S_2} + ... \nonumber \\ & + \frac{1}{a_n} \frac{d'_n h^* + {d'}_n^* h}{S_n} \Bigg] 
\end{align}
we can define the strain data $d_\text{eff}$ for our synthetic coherent detector
\begin{equation}
    d_{\text{eff}} = \left[ \frac{1}{a_1} \frac{d_1}{S_1} + \frac{1}{a_2} \frac{d'_2}{S_2} + ... + \frac{1}{a_n} \frac{d'_n}{S_n} \right] S_{\text{eff}}
\end{equation}
or concisely
\begin{equation}
d_{\text{eff}} = S_{\text{eff}} \sum_n \frac{1}{a_n} \frac{d'_n}{S_n}
\end{equation}
where the power spectral density used to weight inner products is given as
\begin{equation}
    S_{\text{eff}} = \sqrt[n]{S_1 S_2...S_n} \quad \quad \quad S^n_{\text{eff}} = \prod_n S_n
\end{equation}
Now using $\text{E}[(d_n)^2] = S_n(f)$ and no cross terms as before, we derive the power spectral density used to whiten $d_\text{eff}(f)$
\begin{align}
    S^w_{\text{eff}}(f) & = \text{E}[(d_{\text{eff}})^2]  \nonumber \\ & =  \Bigg[ \frac{1}{a_1^2} \frac{\text{E}[(d_1)^2]}{S_1^2} + \frac{1}{a_2^2} \frac{\text{E}[(d'_2)^2]}{S_2^2} + ... \nonumber \\ 
    & \quad \quad \quad \quad \quad \quad \quad  + \frac{1}{a_n^2} \frac{\text{E}[(d'_n)^2]}{S_n^2} \Bigg] S^2_{\text{eff}}
    \nonumber \\
    & = \left[ \frac{1}{a_1^2 S_1} + \frac{1}{a_2^2 S_2} + ... + \frac{1}{a_n^2 S_n} \right] S^2_{\text{eff}}
\end{align}
Thus, 
\begin{equation}
     S^w_{\text{eff}} = S^2_{\text{eff}} \sum_n \frac{1}{a_n^2 S_n}
\end{equation}
which is a Gaussian random variable and whitening $d_\text{eff}(f)$ with $S^w_{\text{eff}}(f)$ preserves Gaussian noise statistics in the synthetic coherent detector.

\subsection{Glitch Rejection Tests}\label{subsec:Methods Glitch Rejection}

Our coherent reconstruction algorithm enforces glitch rejection as two different tests by exploiting the geographic separation of detectors, incoherent time-frequency evolution of glitches, and the preservation of overall Gaussian noise statistics.

According to the Gravity Spy \cite{Gravity_Spy_2017} analyses of Advanced LIGO data for O3 \cite{GWTC-3}, there were $\approx$ 233,981 glitches at Hanford (H1) and $\approx$ 379,805 at Livingston (L1), spanning about 20 distinct glitch morphologies \cite{Glanzer_2023}. Excluding the detector downtimes, and generously assuming that glitches occur at the same rate in every detector, gives us about 1 glitch per minute, or an approximate per detector glitch rate $R_\text{g} \approx 0.017$ Hz. The rate for a pair of incidentally simultaneous glitches in a network of $n$ detectors for a maximum inter-detector light travel time $\Delta t_\text{max} \approx 20$ ms is given by $\lambda_\text{gg} = n R_\text{g}^2 \Delta t_\text{max}$ which corresponds to observing about 1, 1.4, and 2.4 glitches in 24 hours for a network of 2, 3, and 5 detectors, respectively. Thus, more than $99.93\% $ noise transients in the LIGO data are not even incidentally simultaneous and only less than $0.07\% $ need further scrutiny.

\subsubsection{Light travel time test}\label{subsubsection:light travel time test}

The first level of glitch rejection is imposed by the light travel time requirement. Only detectors whose z-statistic time shifts are smaller than the inter-detector light travel time contribute to the coherent synthetic detector. 

Although incidentally simultaneous glitches occur on a daily basis in multi-detector networks, most glitches originate from independent local noise processes, are morphologically distinct, and do not align well using the z-statistic, which measures the cross-correlation between detectors as a function of time shift. Thus, incidentally simultaneous glitches that are morphologically uncorrelated rarely survive this coherence requirement.

For simultaneous glitches that are sufficiently morphologically similar and survive the z-statistic informed light travel time test, we use an additional residual test to flag these "coincident" false alarms.

\subsubsection{Coherent Residuals Test}

To catch false alarms and cases such as coincident glitches, or glitch-signal overlaps, we shift and scale the coherent reconstruction back to the individual detectors $\tilde{h}_\text{coh}^n(f)$ and compute the coherent reconstruction residuals $\tilde{r}_\text{coh}^n(f) = \tilde{d}_n(f) - \tilde{h}_\text{coh}^n(f)$. We re-run the MaxWave glitch model on these residuals to capture any significant incoherent power. As false alarms such as coincident glitches are characterized by independent non-Gaussian structure in each detector, they produce incoherent multi-detector reconstructions. Even if their z-statistic time shift is within the inter-detector light travel time, their distinct time–frequency evolution is reflected in the synthetic detector response, in the resulting coherent reconstruction, and in the residual power after removing the coherent reconstruction.

For signals, $\tilde{h}_\text{resid}(f)$ should either not exist or have occasional low-SNR single wavelets for cases where the initial reconstruction was noisy and the coherent reconstruction removed the noise-induced wavelet. Thus, we flag any $\tilde{h}_\text{resid}(f)$ with more than one wavelet as a "coincident event non-removal" instead of a "signal non-removal". We note that our classification of coincident events or signals is not conclusive, but provisional.

\section{RESULTS}\label{sec:Results}

\begin{figure}
    \centering
    \includegraphics[width=\linewidth]{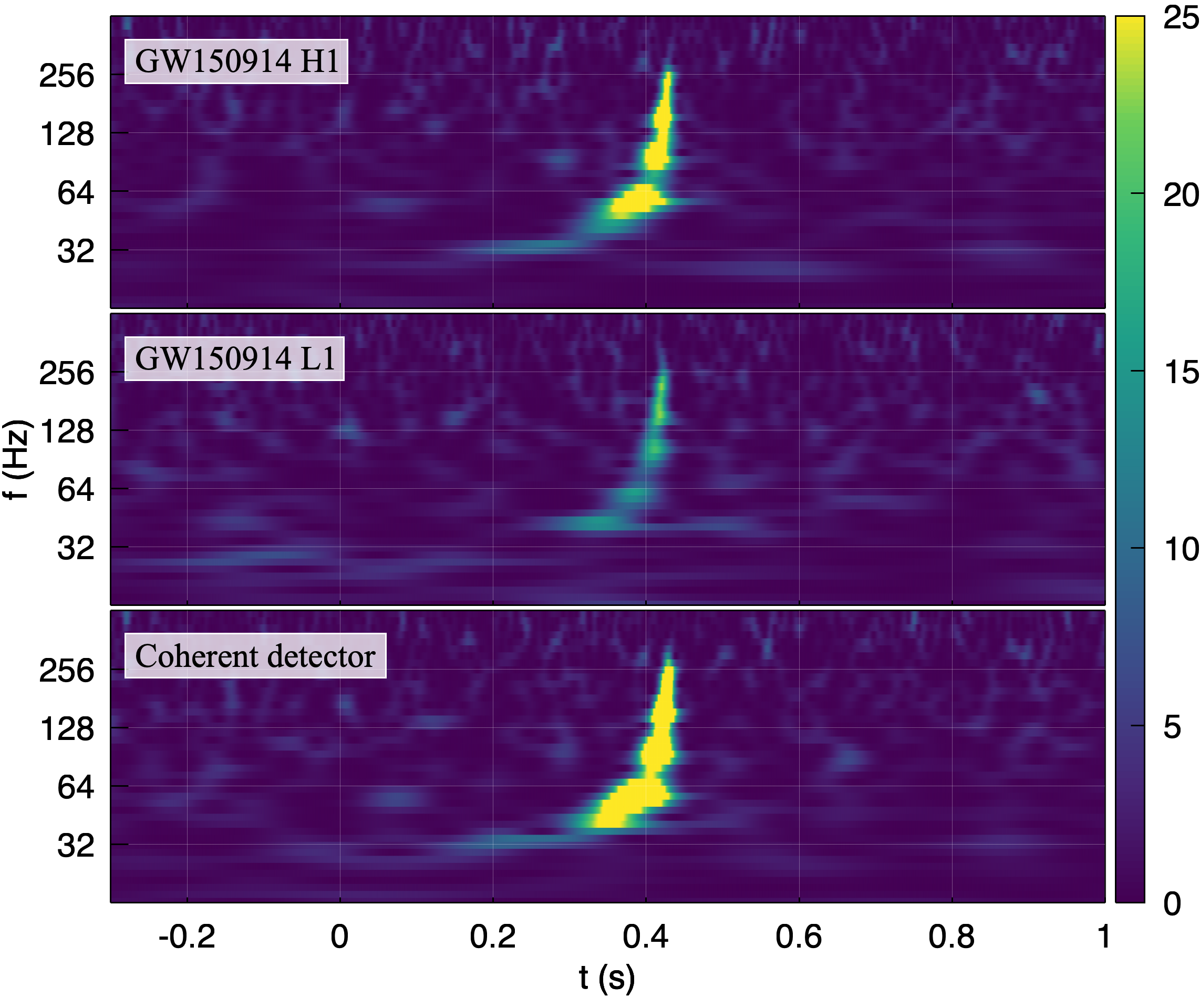}
    \caption{Qscans of the GW150914 event \cite{GW150914, GWTC-1, GWOSC_O1_O3} at $Q=8$ in the individual LIGO detectors H1 (top), L1 (middle), and in our coherent detector (bottom). The coherent detector amplifies the signal while preserving Gaussian noise statistics.}
    \label{fig:GW150914-Qscans}
\end{figure}

We demonstrate improvements in multi-detector signal reconstructions, residuals (Sec. \ref{subsec:GW150914}) and injected signal recovery (Sec. \ref{subsec:Injections}) using MaxWave's coherent detector reconstruction. We compare MaxWave signal model to the \textit{BayesWave} RJMCMC \cite{Cornish_2021} in terms of reconstruction accuracy across various SNRs and binary mass ratios (Sec. \ref{subsec:MW vs BW}). In Sec. \ref{subsec:Results glitch rejection}, we evaluate the performance of our glitch rejection tests against artificially simultaneous glitches. We calculate the operations cost and runtime of the signal model (Sec. \ref{subsection:runtime}) and run it on 14 hours of LIGO data for a two-detector network and estimate our false alarm rate (FAR) using the time-slide method (Sec. \ref{subsec:14hrLIGOrun}). To further test our model on cases where glitches overlap with signals and affect source parameter inference \cite{Udall_2024, Payne_2022, Davis_2022}, we run MaxWave signal on gravitational wave events GW191109 and GW200129 \cite{GWTC-3, GWOSC_O1_O3} (Sec. \ref{subsec:glitches_overlapping_signals}).

\begin{figure}
    \centering
    \includegraphics[width=7.2cm]{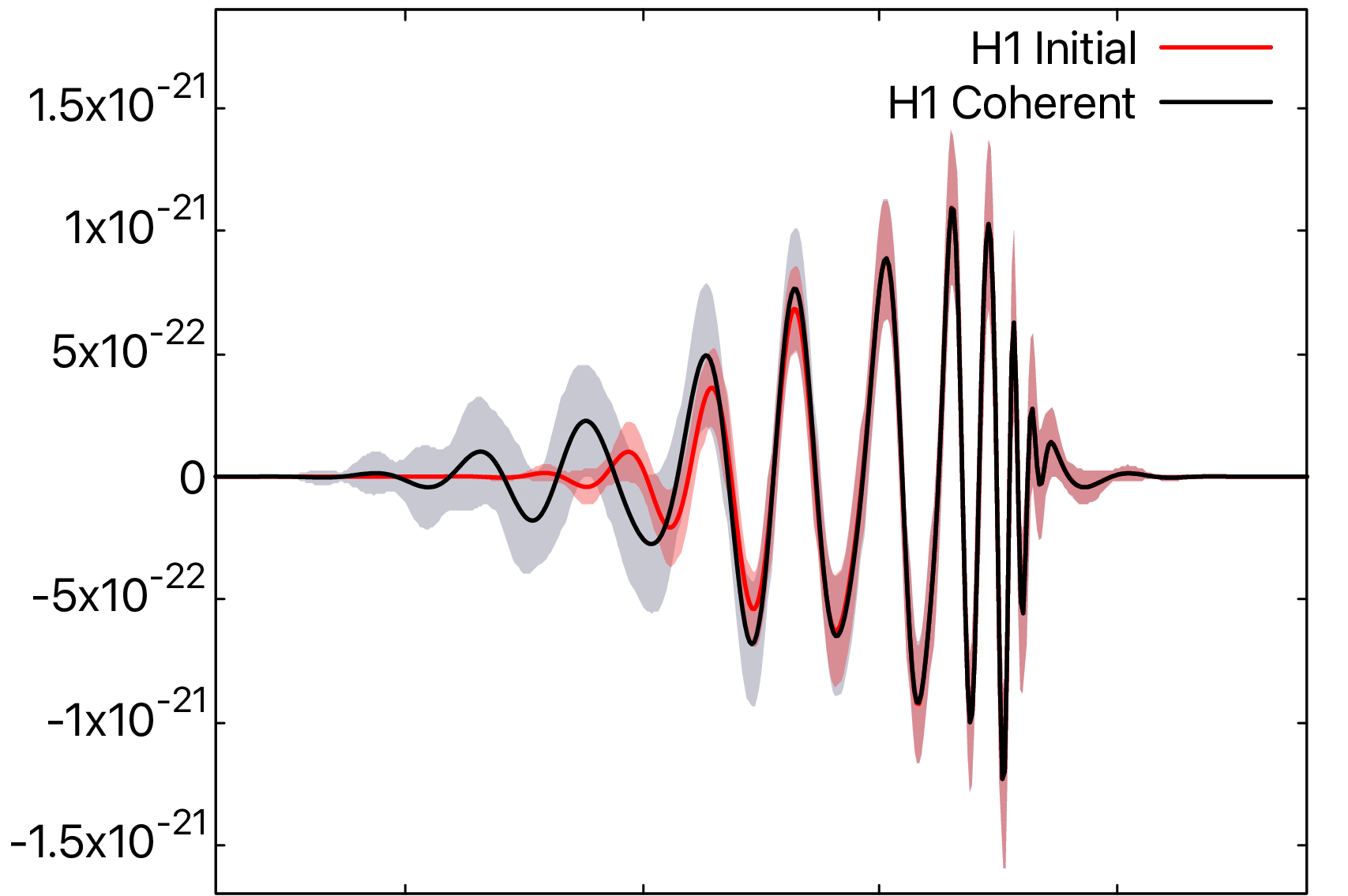}
    \includegraphics[width=7.2cm]{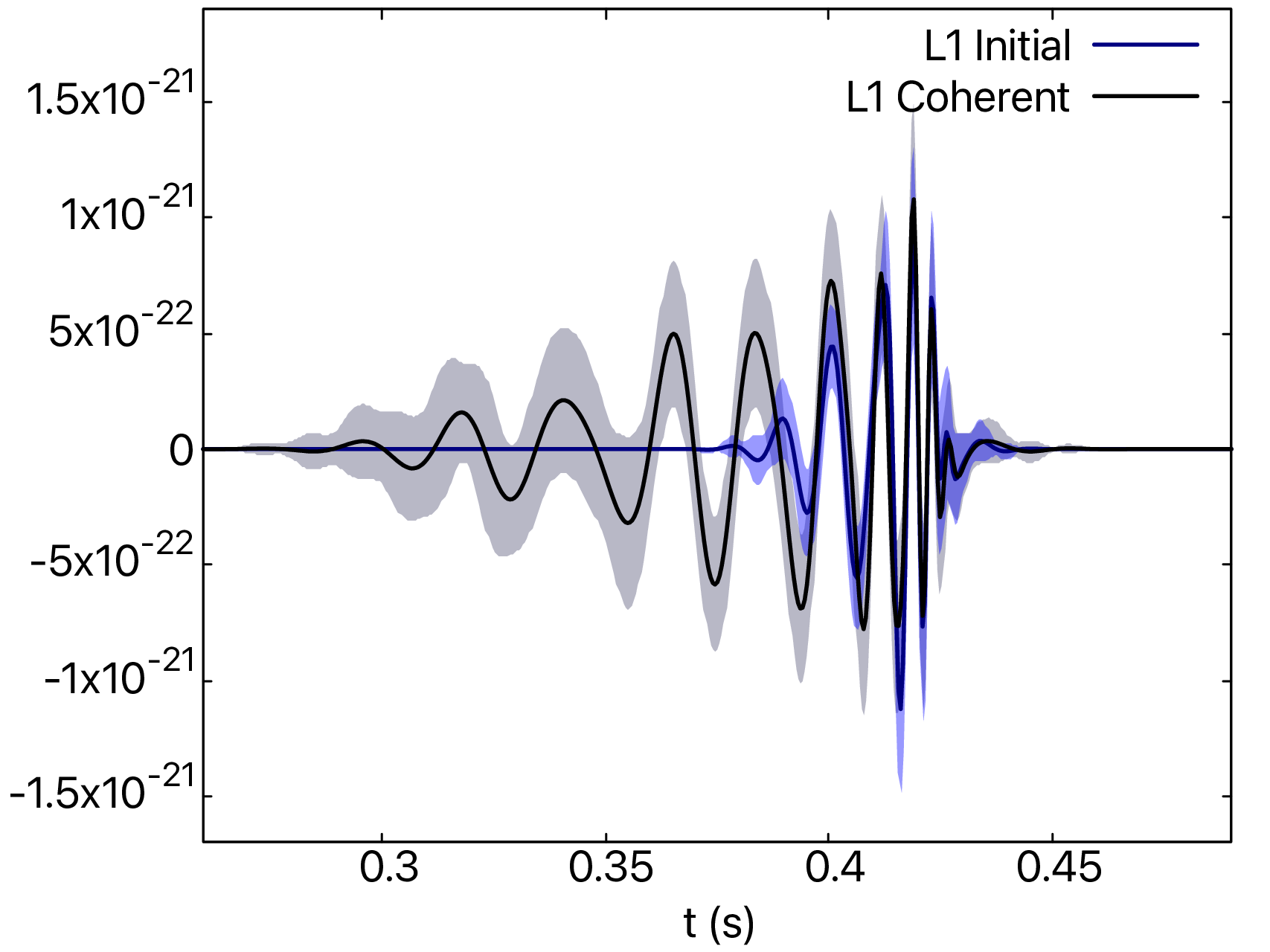}
    \caption{MaxWave's individual and coherent detector time-domain unwhitened reconstructions of the gravitational wave signal GW150914 \cite{GW150914, GWTC-1, GWOSC_O1_O3}. The initial higher-SNR reconstruction for detector H1 (red waveform, top) is aligned with the L1 detector data using z-statistic time and phase shifts, and amplitude scaling to construct the coherent detector as described in Sec. \ref{subsec:comb_multi}. After running MaxWave on the coherent detector, the resulting reconstruction is shifted back to the individual detectors. The shifted coherent reconstructions (black waveforms) render more signal information as compared to the individual H1 (red waveform, top) and L1 (blue waveform, bottom) reconstructions. MaxWave's error envelopes are computed from the Fisher information matrix of the wavelet parameters as described in Sec. II F of \cite{MaxWave_Glitch}.}
    \label{fig:GW150914 reconstructions}
\end{figure}

\subsection{Improvement in signal reconstruction}\label{subsec:GW150914}

\begin{figure*}
    \centering
    \includegraphics[width=\linewidth]{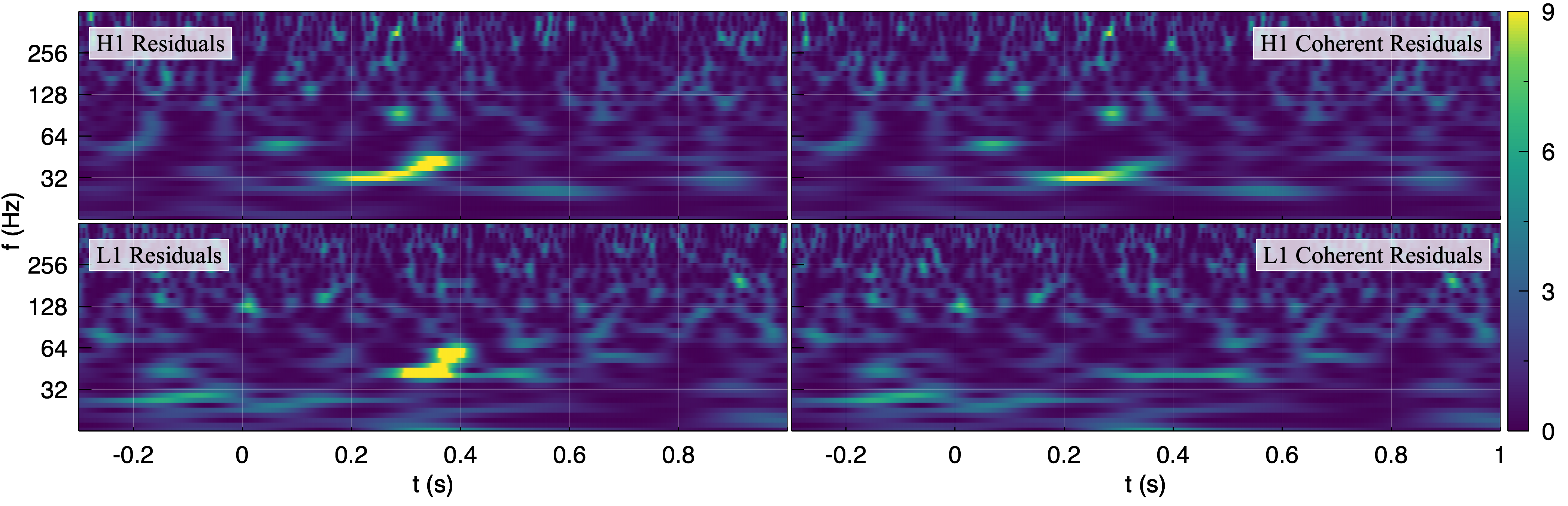}
    \caption{Residuals of the individual (left) and shifted coherent (right) MaxWave reconstructions of the GW150914 event \cite{GW150914, GWTC-1, GWOSC_O1_O3} for H1 (top) and L1 (bottom) detectors. The coherent reconstruction extracts more signal information and produces cleaner residuals.}
    \label{fig:GW150914 residuals}
\end{figure*}

\begin{figure}
    \centering
    \includegraphics[width=\linewidth]{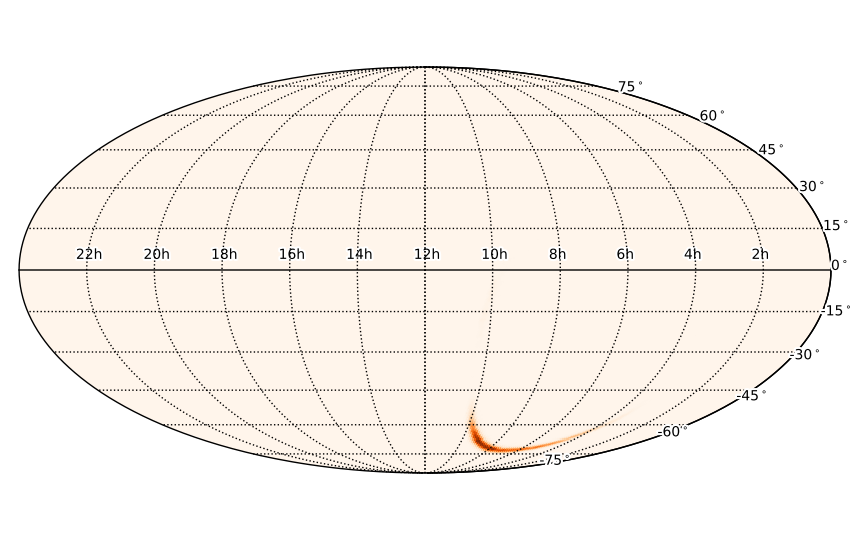}
    \caption{QuickCBC \cite{Neil_QuickCBC} skymap for gravitational wave signal GW150914 \cite{GW150914, GWTC-1, GWOSC_O1_O3} using MaxWave's coherent signal reconstruction as the geocentric template.} \label{fig:skymap_MW_GW150914}
\end{figure}

We present a case study for the gravitational wave event GW150914 \cite{GW150914, GWTC-1, GWOSC_O1_O3} to demonstrate improved coherent detector signal power, feature extraction, and residuals. In Fig. \ref{fig:GW150914-Qscans}, we plot Qscans of the event with quality factor Q $=8$ for the individual LIGO detectors H1 (top), L1 (middle), and for our synthetic coherent detector (bottom). Although MaxWave operates entirely in the time-frequency-time extent (TF$\tau$) regime, Qscans provide better visualizations. The Qscan of the synthetic coherent detector captures the coherent power present in both detectors, amplifying the signal while preserving Gaussian noise statistics.

Fig. \ref{fig:GW150914 reconstructions} shows the individual H1 (red waveform, top) and L1 (blue waveform, bottom) time-domain unwhitened reconstructions of the event. The error envelopes around the reconstructions are computed from the wavelet parameter Fisher information matrix $\Gamma_{kl} = (\partial_k h|\partial_l h)$ as described in Sec. II F of \cite{MaxWave_Glitch}, where detector noise and the finite, approximate parametrization of the model perturb the maximum likelihood solution such that the variance ${\rm Var}[\Delta \lambda^k\Delta \lambda^l] \simeq (\Gamma^{-1})^{kl}$. This gives a per-sample waveform variance ${\rm Var}[h'(t)] = \partial_k h(t) \partial_l h(t) (\Gamma^{-1})^{kl}$ evaluated in the time domain, so that uncertainties in the wavelet parameters and the inverse Fourier transform are both reflected in the envelope. We create the coherent detector as described in Sec. \ref{subsec:comb_multi} by aligning the initial higher-SNR H1 detector reconstruction (red waveform, top) with L1 detector data using z-statistic time and phase shifts, and amplitude scalings. After running MaxWave on the coherent detector, the resulting reconstruction $\tilde{h}_\text{coh}(f)$ is shifted back to each detector (black waveforms). We observe that the shifted coherent reconstruction extracts more signal by combining multi-detector information, especially in the low frequency regime. The match of the reconstructions with the GW150914 numerical relativity waveform template \cite{GW150914, GWTC-1} improves from 93\% for the H1-only reconstruction and 81\% for L1-only reconstruction to 94\% for the shifted coherent reconstructions.

We compare the residuals of the individual (left) and shifted coherent (right) reconstructions for the H1 (top) and L1 (bottom) detectors in Fig. \ref{fig:GW150914 residuals} and, as expected, the shifted coherent reconstructions render better residuals in each detector. In Fig. \ref{fig:skymap_MW_GW150914} we produce a skymap for the event using QuickCBC \cite{Neil_QuickCBC} with the coherent reconstruction as the geocentric template.

\subsection{Injected Signal Recovery}\label{subsec:Injections}

\begin{figure*}
    \centering
    \includegraphics[width=\linewidth]{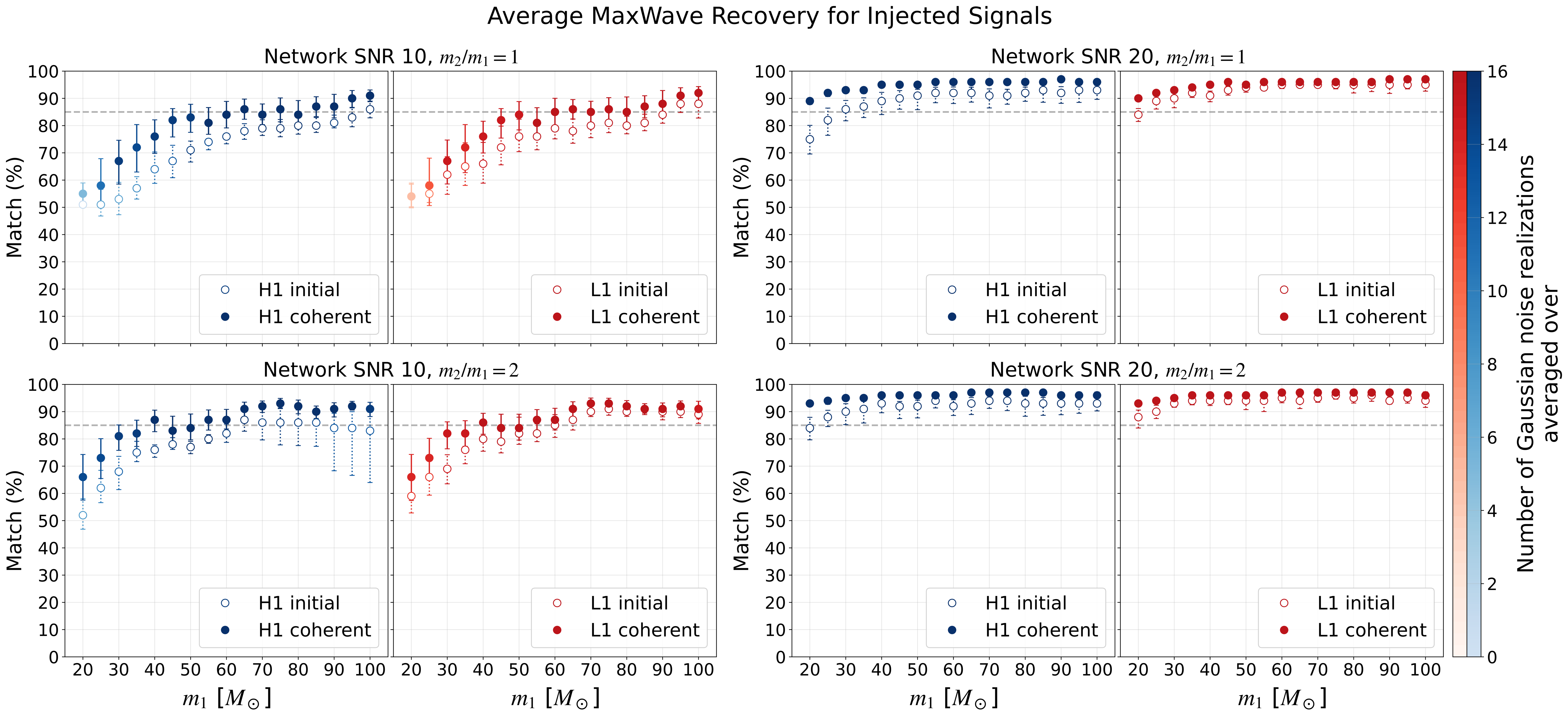}
    \caption{Averaged matches of MaxWave individual and coherent detector recoveries for network SNR 10 (left) and network SNR 20 (right) IMRPhenomD binary black hole injections with $m_1 \in [20, 100] M_\odot$ and mass ratios $m_2/m_1 = 1$ (top) and 2 (bottom). The signals are injected in 16 pairs of 8 s long real LIGO noise realizations from the O3 run \cite{GWTC-3, GWOSC_O1_O3}, cross-assigned between detectors and remapped to a common time axis. For low SNR, MaxWave does not always recover a wavelet. The colorbar indicates an average over the number of times — out of the injections in 16 different LIGO noise realizations — that MaxWave found a wavelet. The blue circles (red circles) indicate averaged match for H1 (L1) single detector recoveries, while the corresponding blue dots (red dots) indicate averaged match for the shifted coherent detector H1 (L1) recoveries. The corresponding error bars give the asymmetrical positive and negative standard deviation around the averaged match. We observe a $6.3_{-1.7}^{+3.3} \%$ improvement in the match for network SNR 10 injections and a $3.2_{-0.7}^{+1.2} \%$ improvement in the match for network SNR 20 injections between the single and coherent detector recoveries.} 
    \label{fig:MW_injection_matches}
\end{figure*}

We compare the matches of individual and coherent MaxWave reconstructions with injected binary black hole waveforms at various SNRs and mass ratios. We inject network SNR 10 and 20 IMRPhenomD binary black hole signals with $m_1 \in [20, 100] M_\odot$ and mass ratios $m_2/m_1=1$ and $2$ in real LIGO noise for a two-detector network of H1 and L1 (Fig. \ref{fig:MW_injection_matches}). The left (right) column plots matches for network SNR 10 (SNR 20) injections. The first (second) row plots matches for mass ratio $m_2/m_1=1$ $(m_2/m_1=2)$. For real LIGO noise, we use 16 pairs of 8 s strain data selected from quiet periods free of gravitational-wave events in the H1 and L1 detectors at various GPS times during the O3 observing runs \cite{GWTC-3, GWOSC_O1_O3}. Segments are cross-assigned between detectors, pairing H1 data from one timestamp with L1 data from another, and remapped to a common time axis. This prevents the two detectors from sharing a GPS epoch, eliminating any accidental low-SNR signal from contaminating the noise. 

\begin{figure*}
    \centering
    \includegraphics[width=\linewidth]{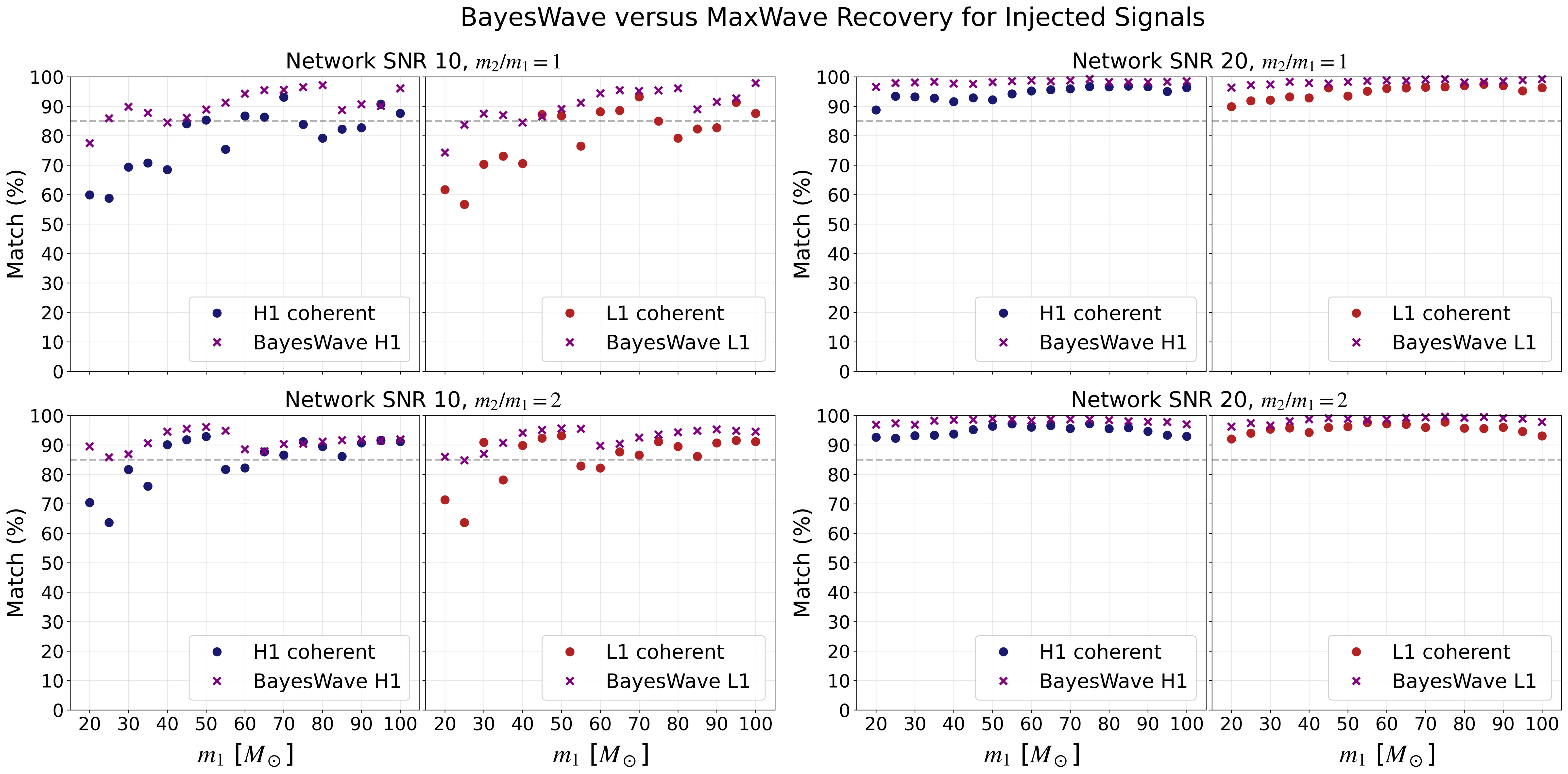}
    \caption{Comparing matches of \textit{BayesWave} two detector \cite{Cornish_2021} (crosses) and MaxWave signal model (dots) recoveries for network SNR 10 (left) and network SNR 20 (right) IMRPhenomD binary black hole injections with $m_1 \in [20, 100] M_\odot$ and mass ratios $m_2/m_1 =1$ (top) and 2 (bottom) in 8 s of real LIGO noise at GPS trigger times 1264213223 for detector H1 and 1266238142 for detector L1 \cite{GWTC-3, GWOSC_O1_O3}. Strain from both detectors is remapped to a common time axis before injecting signals. Compared to the robust but computationally intensive \textit{BayesWave}, MaxWave may trade reconstruction accuracy to provide a real-time, low-latency, approximate maximum likelihood solution. However, as we increase SNR and individual binary masses, recovered waveforms matches for the MaxWave signal model converge towards the \textit{BayesWave} recovery. The dips in the match with increasing $m_1$ show us that both \textit{BayesWave} and MaxWave are sensitive to how the injected waveform aligns with specific features of the noise segment \cite{MaxWave_Glitch, Cornish_2021}. These fluctuations disappeared when we averaged over MaxWave matches for signals injected in 16 real LIGO noise realizations in Fig. \ref{fig:MW_injection_matches}. Similar to \textit{BayesWave}, MaxWave performs better for higher-mass binaries as they occupy smaller time–frequency volumes and localize power within a few higher-SNR wavelets \cite{MaxWave_Glitch, Cornish_2021}.}
    \label{fig:BW_versus_MW_injections}
\end{figure*}

For low SNR, MaxWave does not always recover a wavelet. The colorbar indicates an average over the number of times — out of the injections in 16 different LIGO noise realizations — that MaxWave found a wavelet. The blue circles (red circles) indicate averaged match for H1 (L1) single detector recoveries, while the corresponding blue dots (red dots) indicate averaged match for the shifted coherent detector H1 (L1) recoveries. The corresponding error bars give the asymmetrical positive and negative standard deviation around the averaged match. For coherent detector recoveries, we observe a $6.3_{-1.7}^{+3.3} \%$ improvement in the match for network SNR 10 injections and a $3.2_{-0.7}^{+1.2} \%$ improvement in the match for network SNR 20 injections as compared to the single detector recoveries.

We can further maximize the log likelihood of the shifted coherent reconstructions in each detector through the refined MaxWave glitch model, which perturbs the wavelet parameters $t, f,$ and $ \tau$ beyond their fixed grid (described in Sec. IIG of \cite{MaxWave_Glitch}). Maximizing log L can lead us to fit more noise and does not always
improve the matches of the recovered waveforms with the
corresponding injected templates, especially for low SNR signals. Refining the coherent detector reconstruction in each detector only marginally further improves the matches by $0.6_{-0.9}^{+0.5} \%$ for network SNR 10 and $0.4_{-0.4}^{+0.2} \%$ for network SNR 20 injections.

\subsection[\textit{BayesWave}]{MaxWave versus \textit{BayesWave} RJMCMC}\label{subsec:MW vs BW}

In Fig. \ref{fig:BW_versus_MW_injections}, we compare the match between the injected and recovered waveforms for the \textit{BayesWave} \cite{Cornish_2021} two detector (crosses) and the MaxWave coherent signal model (dots) reconstructions. We inject network SNR 10 and 20 IMRPhenomD binary black hole signals with $m_1 \in [20, 100] M_\odot$ and mass ratios $m_2/m_1=1$ and $2$ in 8 s of real LIGO noise at GPS trigger times 1264213223 for detector H1 and 1266238142 for detector L1 \cite{GWTC-3, GWOSC_O1_O3}. Strain from both detectors is remapped to a common time axis before injecting signals. The left (right) column plots matches for network SNR 10 (SNR 20) injections. The first (second) row plots matches for mass ratio $m_2/m_1=1$ $(m_2/m_1=2)$. We expect the matches between the injected and recovered waveforms for the MaxWave signal model (dots) to be lower than those for \textit{BayesWave} (crosses). While \textit{BayesWave} reconstructs signals through a computationally intensive RJMCMC with trans-dimensional sampling and requires hours to run, MaxWave provides a real-time, low-latency, approximate maximum likelihood signal model. Consequently, MaxWave may trade reconstruction accuracy for substantial computational efficiency. However, as we increase SNR and individual binary masses, recovered waveforms matches for the MaxWave coherent signal model converge towards the \textit{BayesWave} recovery. The dips in the match with increasing $m_1$ in Fig. \ref{fig:BW_versus_MW_injections} are incidental. These variations show us that both \textit{BayesWave} and MaxWave are sensitive to how the injected waveform aligns with specific features of the noise segment. Small changes in waveform morphology can interact differently with the noise, leading to fluctuations in the recovered match \cite{MaxWave_Glitch, Cornish_2021}. These fluctuations disappeared when we averaged over MaxWave matches for signals injected in the 16 pairs of real LIGO noise realizations (Fig. \ref{fig:MW_injection_matches}). 

At fixed SNR, both \textit{BayesWave} (crosses in Fig. \ref{fig:BW_versus_MW_injections}) and the MaxWave signal model (dots in Fig. \ref{fig:MW_injection_matches}) perform better for higher-mass binaries, which occupy smaller time–frequency volumes, localize power within a few higher-SNR wavelets and are more efficiently reconstructed. On the other hand, lower mass binaries span many cycles across a wide frequency range, distribute SNR over a large time–frequency volume, increase the number of low-SNR wavelets needed to reconstruct the signal, and reduce reconstruction fidelity \cite{MaxWave_Glitch}. This behavior is consistent with the known performance of \textit{BayesWave} \cite{Cornish_2021}. Any improvement in recovery with increasing total mass is limited by the detector noise spectrum, as extremely massive systems below the sensitive frequency band are less recoverable.

\subsection{Glitch rejection}\label{subsec:Results glitch rejection}

\begin{figure}[b]
    \centering
    \includegraphics[width=\linewidth]{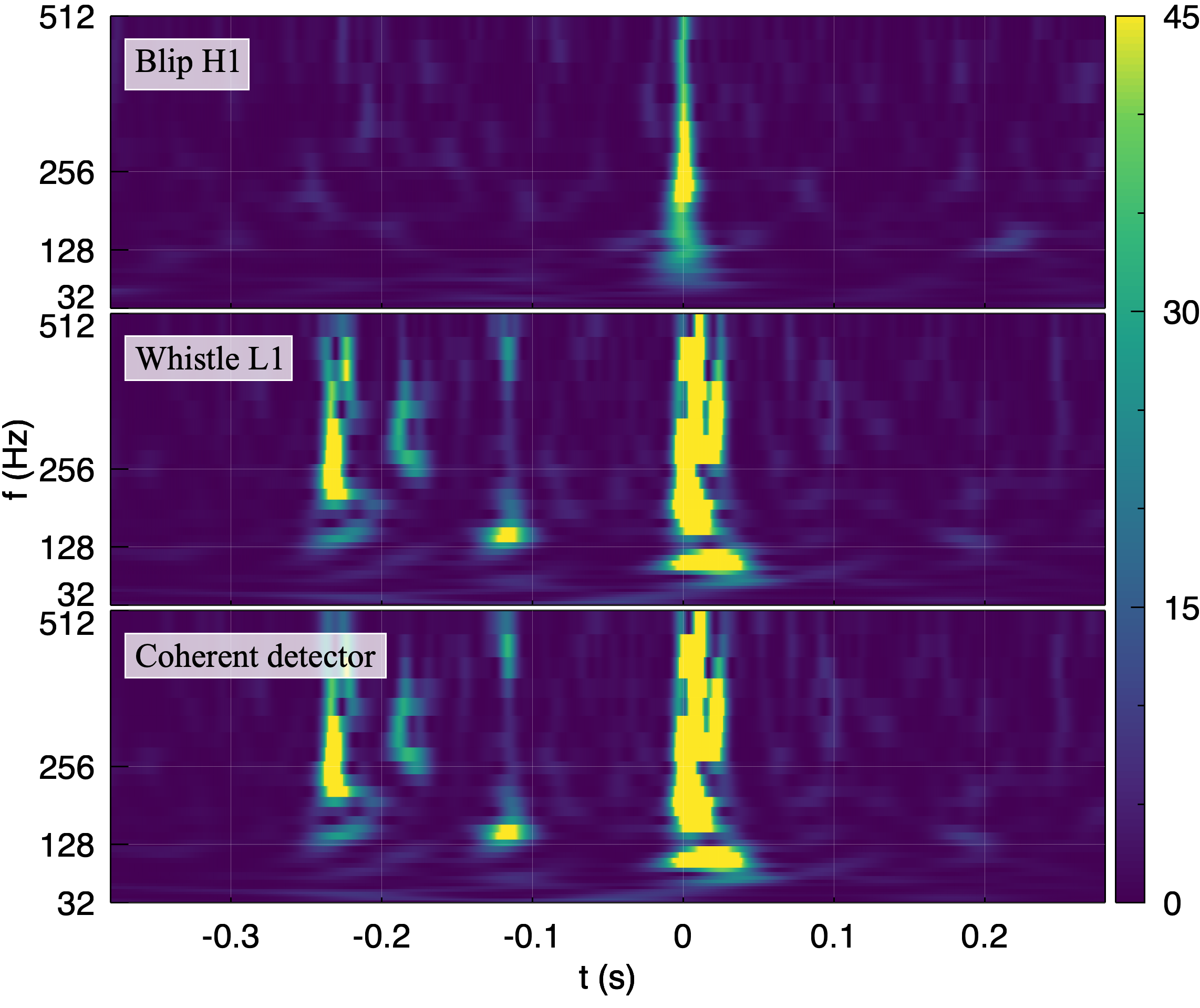}
    \caption{Blip in H1 at GPS time 1168989748.12647 (top) and Whistle in L1 (middle) at GPS time 1253426470.4976 \cite{GWTC-2, GWTC-3, GWOSC_O1_O3} remapped to a common time axis and artificially placed within the inter-detector light travel time window. The coherent detector (bottom) aligns the glitches using z-statistic time and phase shifts, and amplitude scalings, amplifying any coherent power while maintaining Gaussian noise statistics.}
    \label{fig:glitch-glitch Qscan}
\end{figure}

\begin{figure*}
    \centering
    \includegraphics[width=\linewidth]{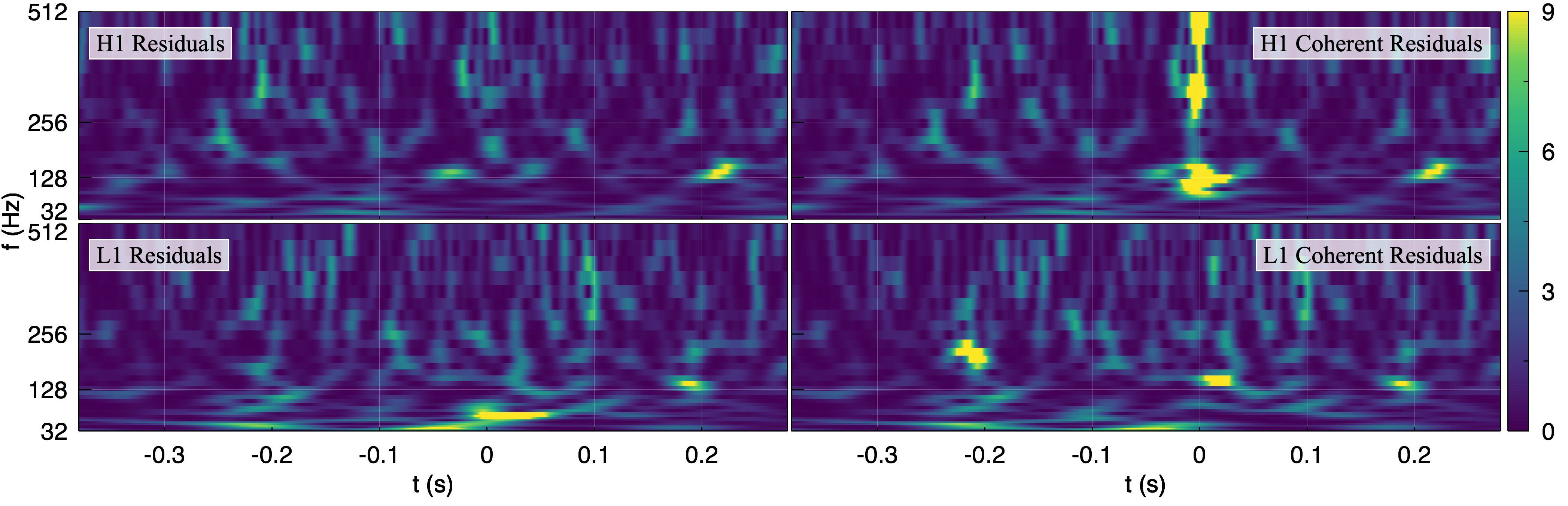}
    \caption{Individual (left) and coherent (right) reconstruction residuals for Blip in H1 (top) and Whistle in L1 (bottom). Glitches are morphologically distinct and their coherent residuals contain remnant power. Re-running MaxWave on the coherent residuals produces $\tilde{h}_\text{resid}^n(f)$ with NW $> 1$, fails the clean residuals test and correctly flags the case as "coincident event non-removal".}
    \label{fig:glitch-glitch residuals}
\end{figure*}

As incidentally simultaneous glitches are rare, we artificially construct cases where glitches occur within the light travel time to further test our flagging and rejection criteria. Any event that passes the z-statistic based light travel time test with sufficient SNR in at least two detectors is marked as a "non-removal". The coherent residuals test then only provisionally flags these cases as "signal non-removal" or "coincident event non-removal".

We demonstrate the coherent residual test for a blip glitch in the H1 detector with SNR = 23.9 originally at GPS time 1168989748.12647 and a whistle glitch in the L1 detector with SNR = 77.5 originally at GPS time 1253426470.4976 \cite{GWTC-2, GWTC-3, GWOSC_O1_O3} by remapping the glitches to a common time axis and artificially placing them within the inter-detector light travel time window. Fig. \ref{fig:glitch-glitch Qscan} shows the Qscans for the individual detectors and the coherent detector, where H1 contains the blip (top) and L1 the whistle (middle). The coherent detector aligns these glitches using the z-statistic time and phase shifts and amplitude scalings, amplifying any coherent power (bottom). However, as the glitches are morphologically distinct in time-frequency space, they are inherently incoherent and leave behind residual power after subtracting the coherent reconstruction. We can see this remnant power in the residuals for the coherent H1 (right, top) and coherent L1 (right, bottom) reconstructions  in Fig. \ref{fig:glitch-glitch residuals}, which are significantly more contaminated than the residuals for the individual H1 (left, top) and L1 (left, bottom) detector reconstructions. Re-running MaxWave on the coherent residuals reconstructs $\tilde{h}_\text{resid}^n(f)$ for both the H1 and L1 with two wavelets and SNRs 11.0 and 7.1 respectively, failing our clean residuals test for the "signal non-removal" flag. Instead, this case is accurately flagged as "coincident event non-removal".

To further quantify simultaneous glitch rejection, we construct simultaneous glitches using 10 categories of 99.99\% confident Gravity Spy \cite{Gravity_Spy_2017} glitches, including 60 glitches per category for blips, low-frequency blips, whistle, koi fish, tomte, fast scattering, scattered light, scratchy, and low-frequency bursts, and 56 extremely loud glitches. These glitches are randomly sampled across the LIGO O3 run \cite{GWTC-3, GWOSC_O1_O3} from H1 and L1 detectors, mapped to a common time axis, and artificially placed within the inter-detector light travel time. MaxWave signal's z-statistic light travel time test alone rejects $74.4\%$ of the 2,960 simultaneous glitches analyzed, and the coherent residuals test increases the rejection rate to $94.0\%$. Thus, the probability of incorrectly flagging coincident glitches as signals is about $6.0\%$. As expected, cross-type glitch coincidences have a much higher rejection rate of 95.9\% than that of same-type glitches at 76.0\%. The z-statistic test alone struggles to reject simultaneous blips, low-frequency blips, tomte, koi fish, and extremely loud pairs as glitches in these categories can be morphologically similar. The coherent residual test also particularly struggles for simultaneous blip, low-frequency blip, and tomte pairs.

For our initial estimate of observing about 1 incidentally simultaneous glitch in 24 hours for a network of 2 detectors, our false alarm rate for simultaneous glitches incorrectly flagged as signals is less than about 1 in 16 days or $6.9 \times 10^{-7}$ Hz. For the H1-L1 network specifically, simultaneous glitches occur about once in four days, and this false alarm rate further drops to about 1 in two months or $1.7 \times 10^{-7}$ Hz.

\subsection{Operations Cost and Runtime}\label{subsection:runtime}

The z-statistic alignment and additional MaxWave runs contribute to the operations cost and runtime increase between the MaxWave glitch \cite{MaxWave_Glitch} and signal models. The z-statistic alignment computes the phase-blind cross-correlation between up to $n$ frequency-domain waveforms via two IFFTs of length $2 N$, where $N$ is the number of time domain samples and the factor of two arises from zero-padding. Thus, the additional computational cost from the multi-detector alignment is up to $(n-1) \mathcal{O}(2N \log{2N}) \approx 2(n-1)\mathcal{O}(N \log{N})$. For the cases that pass the z-statistic light travel time test, MaxWave glitch is run $2n+1$ times in total, including the $n$ initial runs per individual detector, a run on the coherent detector, and $n$ residual runs for the individual detectors after subtracting the shifted coherent reconstructions. The noise spectrum is estimated once per detector in the initial $n$ runs.

For $T_{\text{obs}} = 4$ s and $N_\tau = 6$ time extent layers in the glitch model's wavelet transform (Table 1. in \cite{MaxWave_Glitch}), the MaxWave signal model runs in about $0.8$ to $1.32$ s for two detectors, $1.16$ to $1.94$ s for three detectors, and $1.88$ to $3.18$ s for five detectors, depending on whether the shifted coherent reconstructions are refined $50$ times or not (Sec. II G in \cite{MaxWave_Glitch}). Even with $N_\tau = 8$, the signal model runs in about $0.86$ to $1.4$ s for two detectors, $1.25$ to $2.06$ s for three detectors, and $2.03$ to $3.38$ s for five detectors, where all CPU runtimes are measured on a single core
Intel i7-7820HQ (2.9 GHz) processor. Thus, MaxWave signal model, with up to 50 perturbative refinements to the shifted coherent reconstructions in each detector, maintains real time performance for a five-detector network. 

\subsection{Real LIGO data analysis}\label{subsec:14hrLIGOrun}
\begin{figure*}
    \centering
    \includegraphics[width=\linewidth]{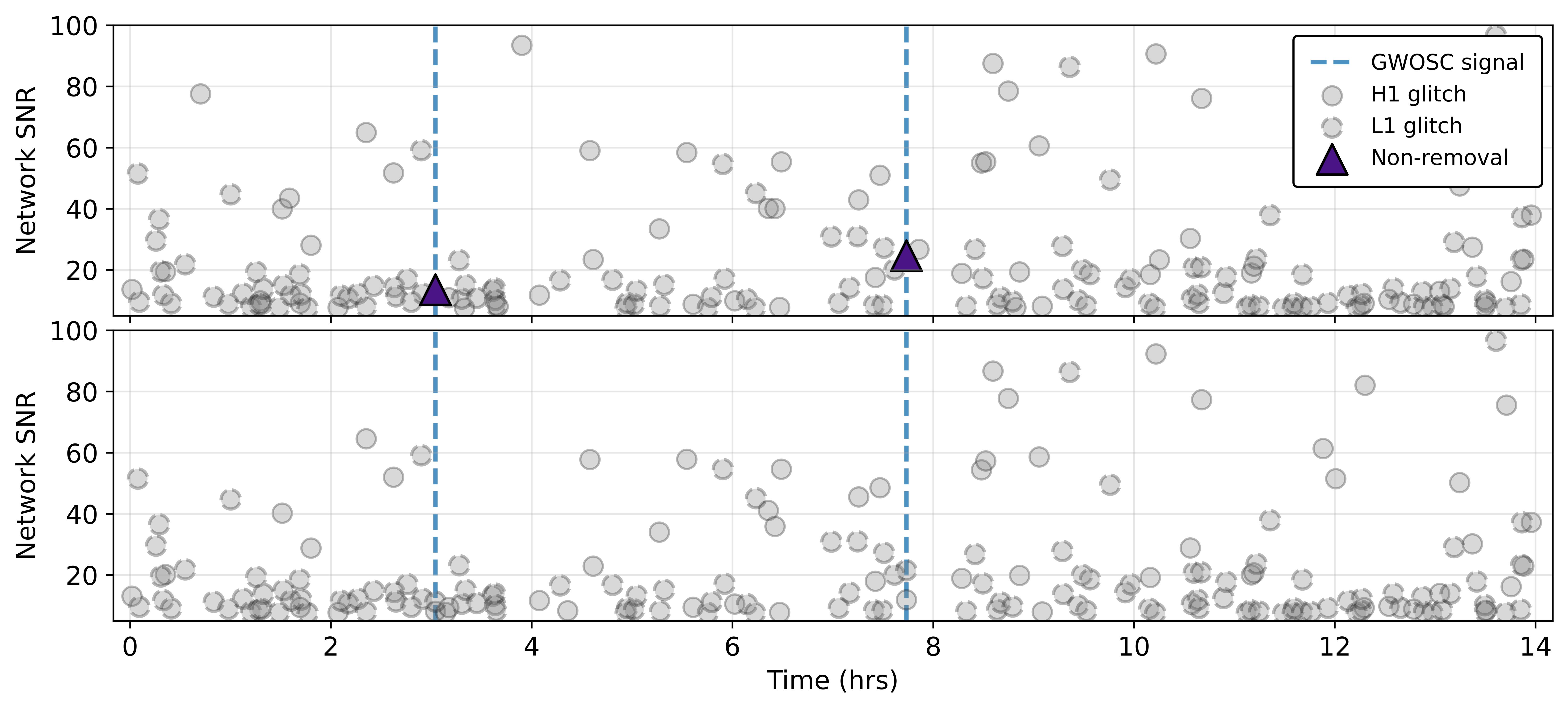}
    \caption{MaxWave signal model run on 14 hours of LIGO data on May 21, 2019 for a two-detector network with H1 and L1 detectors (top panel) \cite{GWTC-3} and on time-slid data with L1 detector strain shifted by 1 s relative to the H1 strain (bottom panel). The signal non-removals (purple triangles) recovered by the model with SNRs 14.8 and 25.1 in the top panel are consistent with gravitational wave events GW190521\_030229 with SNR $=14.3^{+0.5}_{-0.4}$ at GPS time 1242442967.4 and GW190521\_074359 with SNR $=25.9^{+0.1}_{-0.2}$ at GPS time 1242459857.4 in the GWOSC event catalog \cite{GWOSC_O1_O3}. The 14 hour time-slid run (bottom panel) produces no false alarms. Thus, MaxWave signal model's FAR $\leq 1.99 \times10^{-5}$ Hz, which corresponds to less than two false alarms in 24 hours.}
    \label{fig:frames and frames_slid}
\end{figure*}

We run the MaxWave signal model on 14 hours of LIGO data on May 21, 2019 for a two detector network with H1 and L1 detectors \cite{GWTC-3}. The top panel of Fig. \ref{fig:frames and frames_slid} shows the "signal non-removals" (purple triangles), H1 glitches (gray circles, solid border) and L1 glitches (gray circles, dotted border) that MaxWave reconstructs. The signal non-removals flagged by MaxWave are consistent with the gravitational wave events GW190521\_030229 at GPS time 1242442967.4 and GW190521\_074359 at GPS time 1242459857.4 recorded in the Gravitational Wave Open Science Center (GWOSC) catalog (dashed vertical lines) \cite{GWOSC_O1_O3}. MaxWave signal model SNRs for these events are 14.8 and 25.1 respectively, which are comparable to the SNRs $14.3^{+0.5}_{-0.4}$ and $25.9^{+0.1}_{-0.2}$ in the GWOSC catalog.

\begin{figure}
    \centering
    \includegraphics[width=\linewidth]{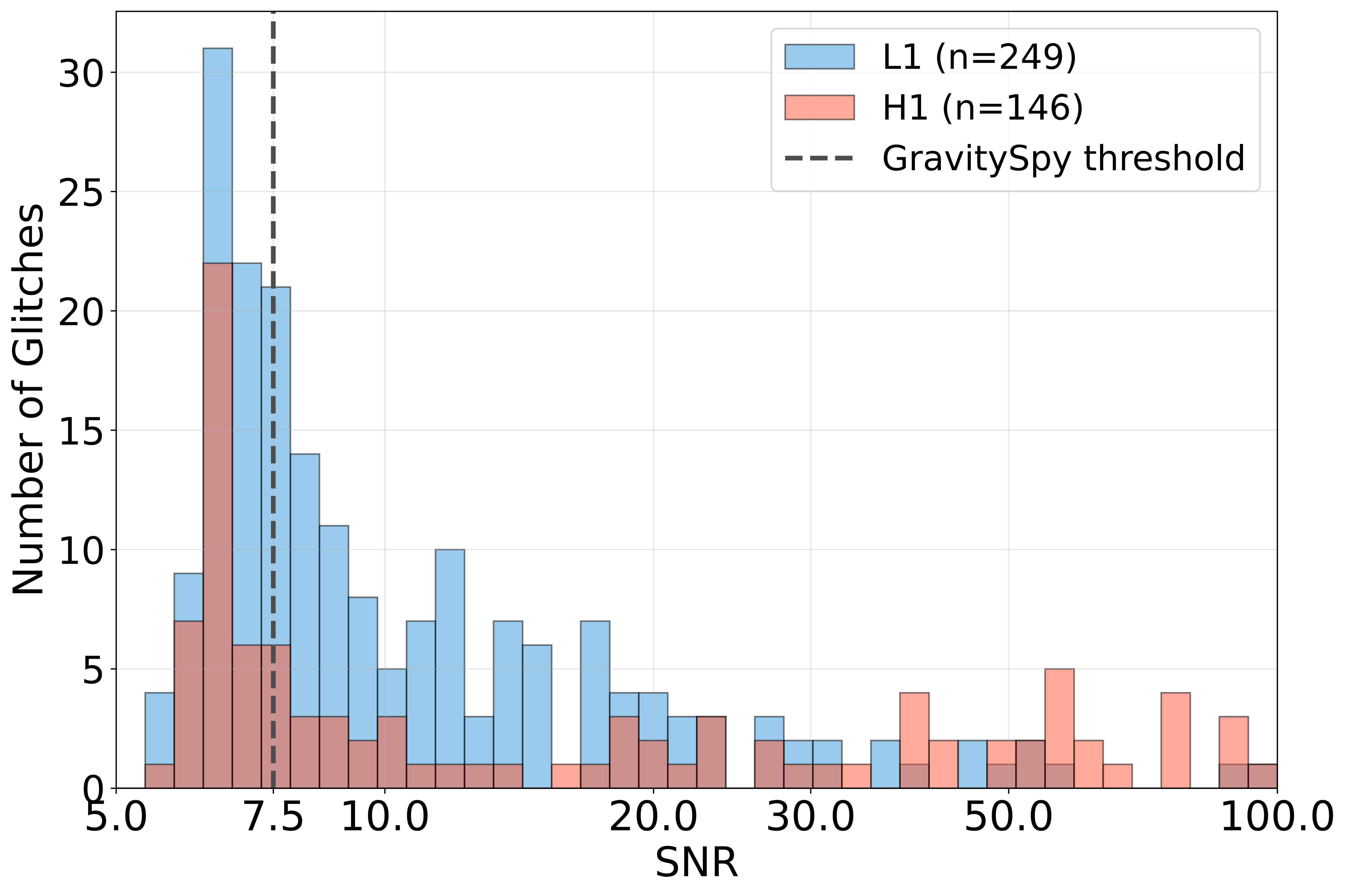}
    \caption{Histograms of the H1 and L1 detector glitches with SNR $\in [5,100]$ reconstructed by the MaxWave signal model run on 14 hours of LIGO data from May 21, 2019 \cite{GWTC-3}. MaxWave signal reconstructs glitches at lower SNRs than the SNR 7.5 Gravity Spy threshold \cite{Gravity_Spy_2017}.}
    \label{fig:frame_hist}
\end{figure}

To compute the FAR of our model, we use the time-slide method and shift the 14 hour L1 data by 1 s with respect to the H1 data, de-correlating true gravitational wave signals across the network. The bottom panel of Fig. \ref{fig:frames and frames_slid} highlights that MaxWave only reconstructs glitches in the time-slid case and produces no false alarms for the 14 hour time-slid data. Thus, MaxWave signal model's FAR $\leq 1.99 \times10^{-5}$ Hz, which corresponds to less than two false alarms in 24 hours.

\begin{figure*}
    \centering
    \includegraphics[width=4.4cm]{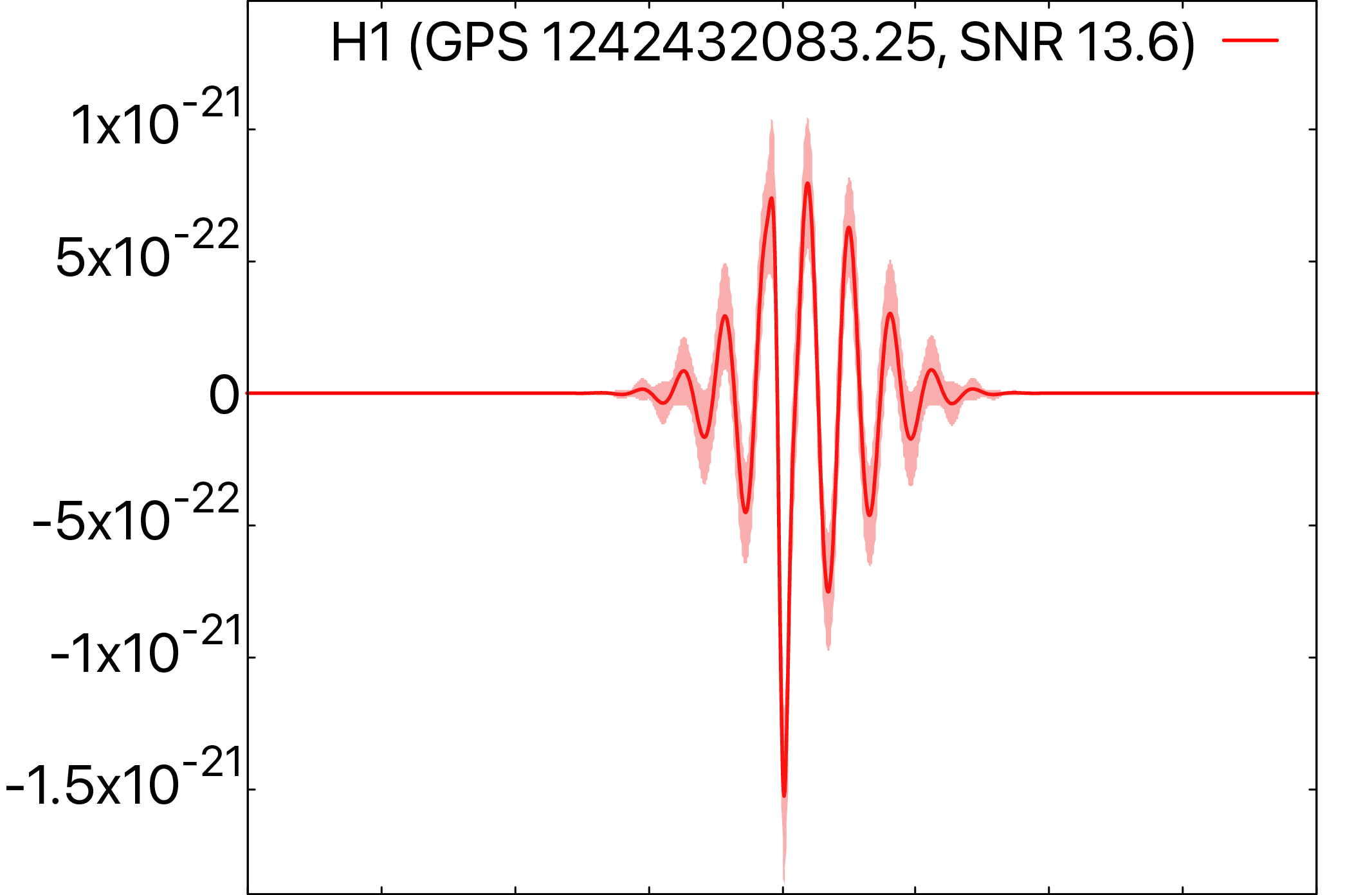}%
    \includegraphics[width=4.4cm]{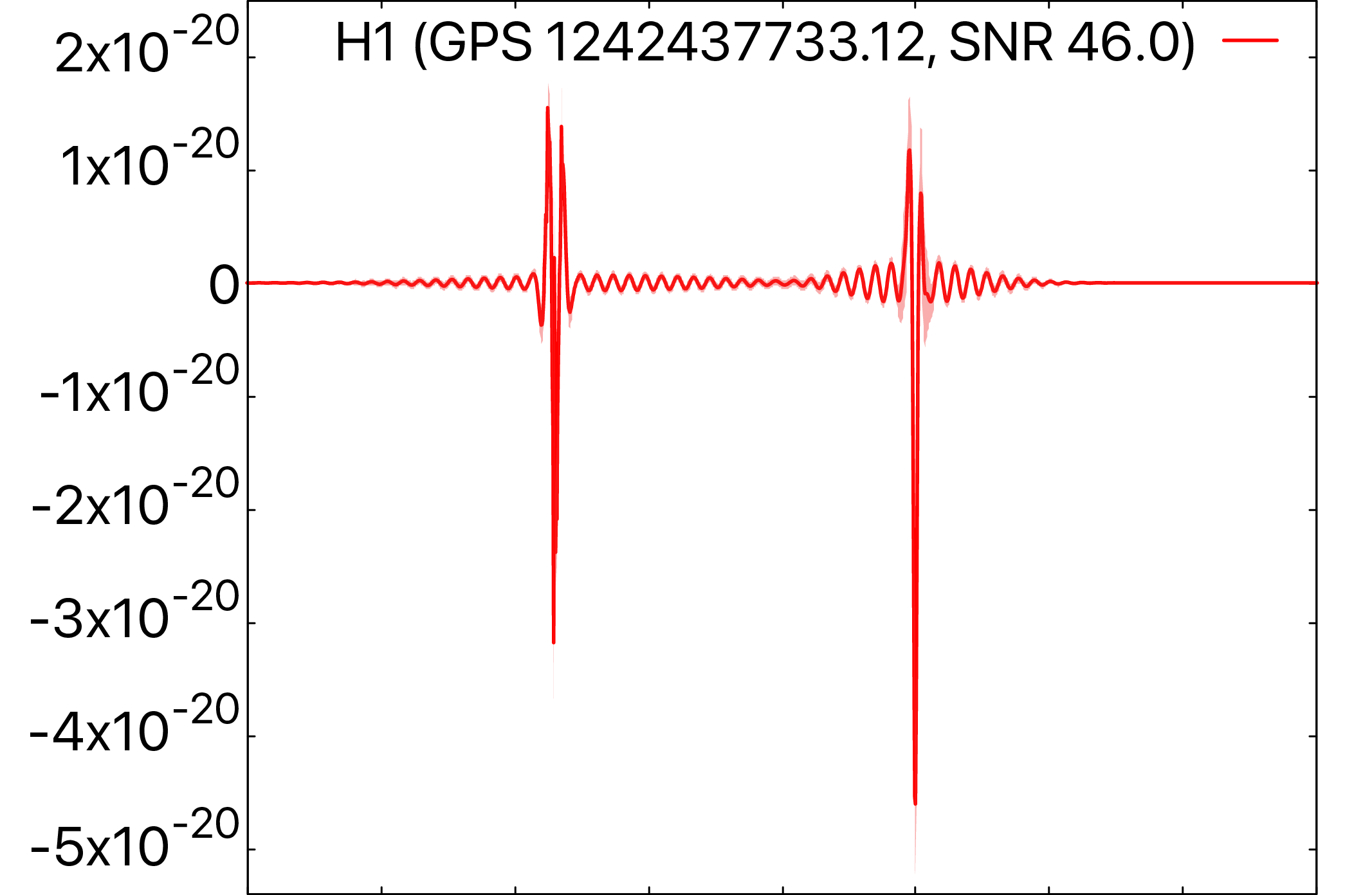}%
    \includegraphics[width=4.4cm]{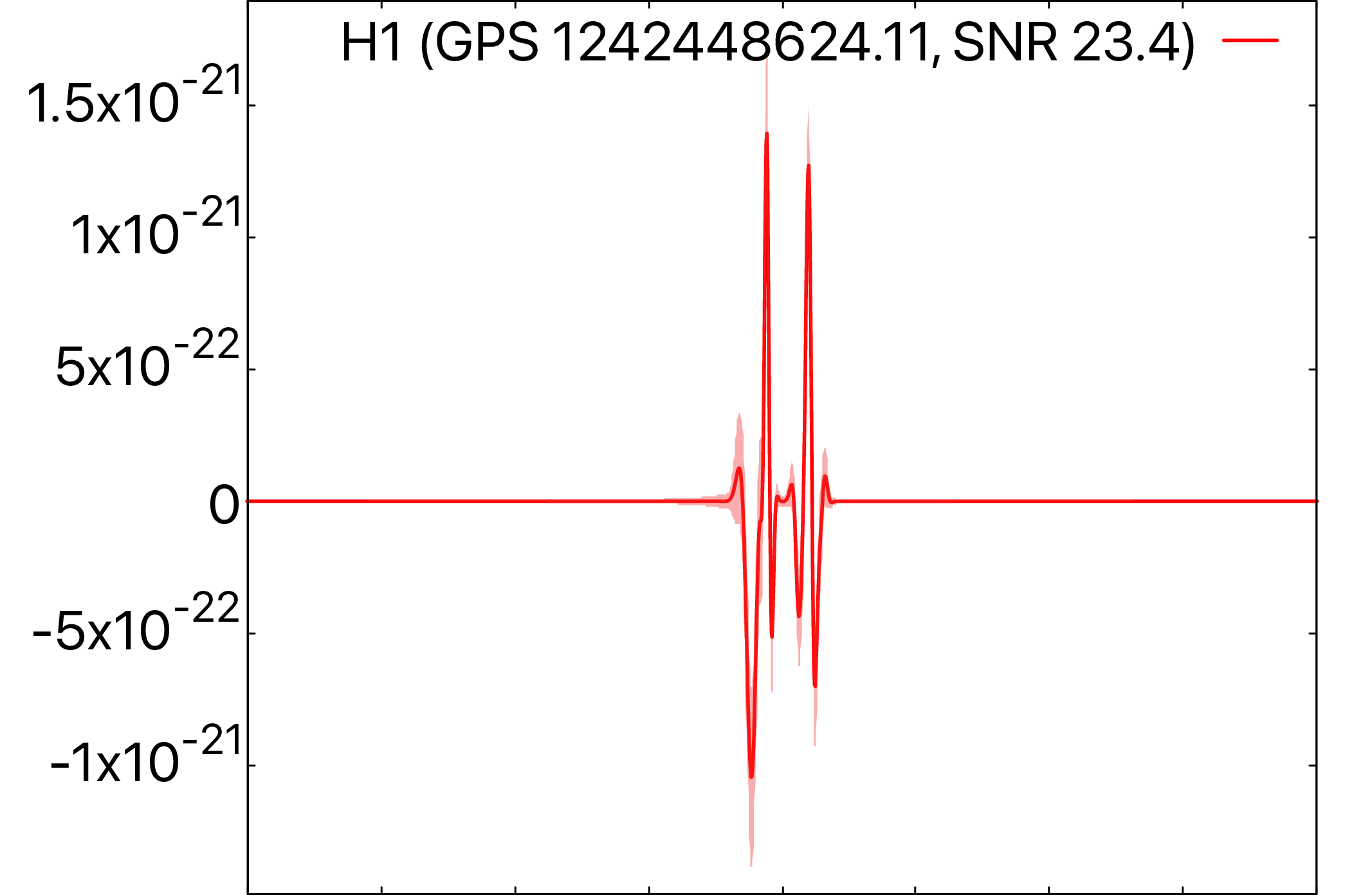}%
    \includegraphics[width=4.4cm]{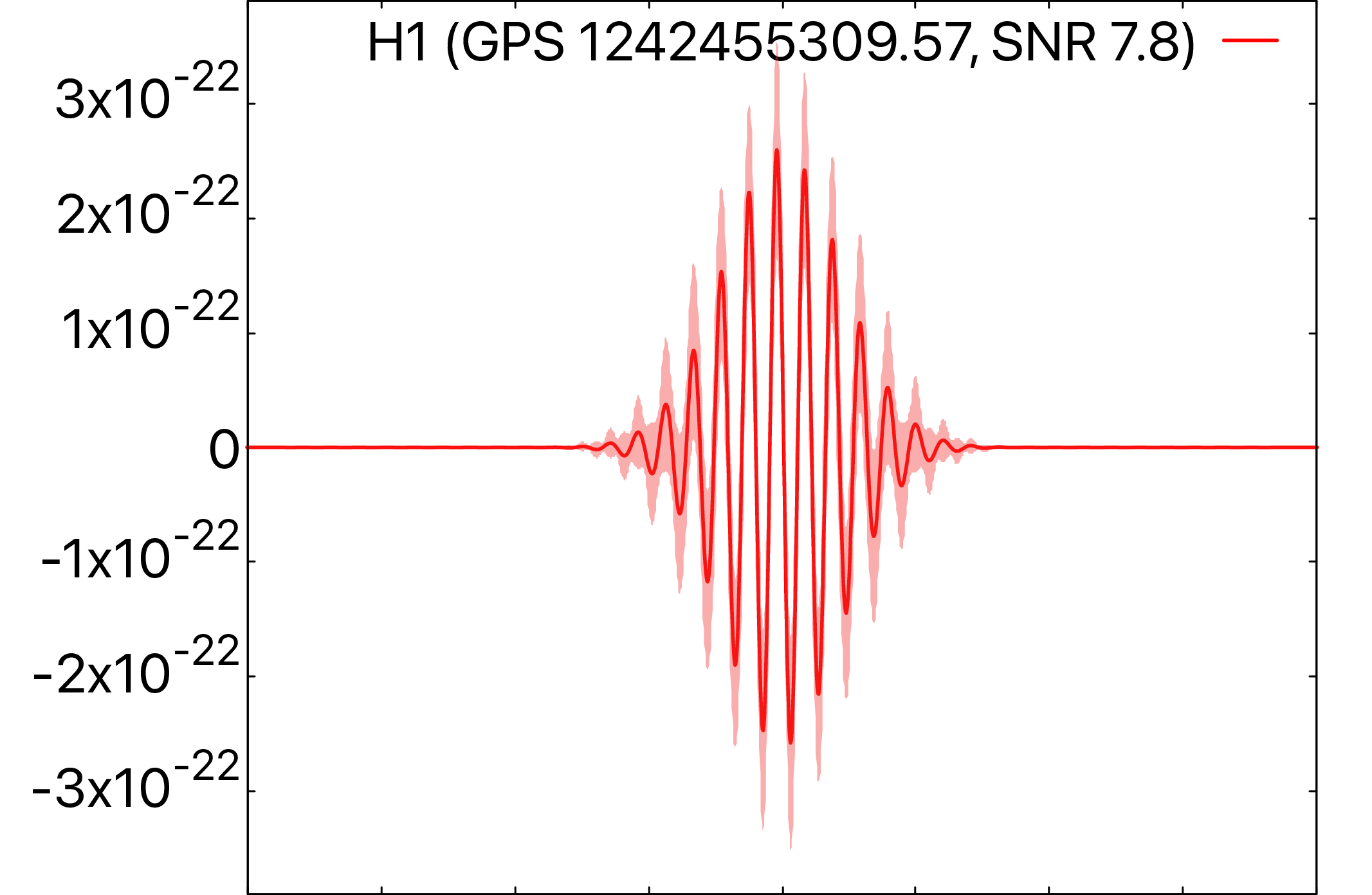}\\
    \includegraphics[width=4.4cm]{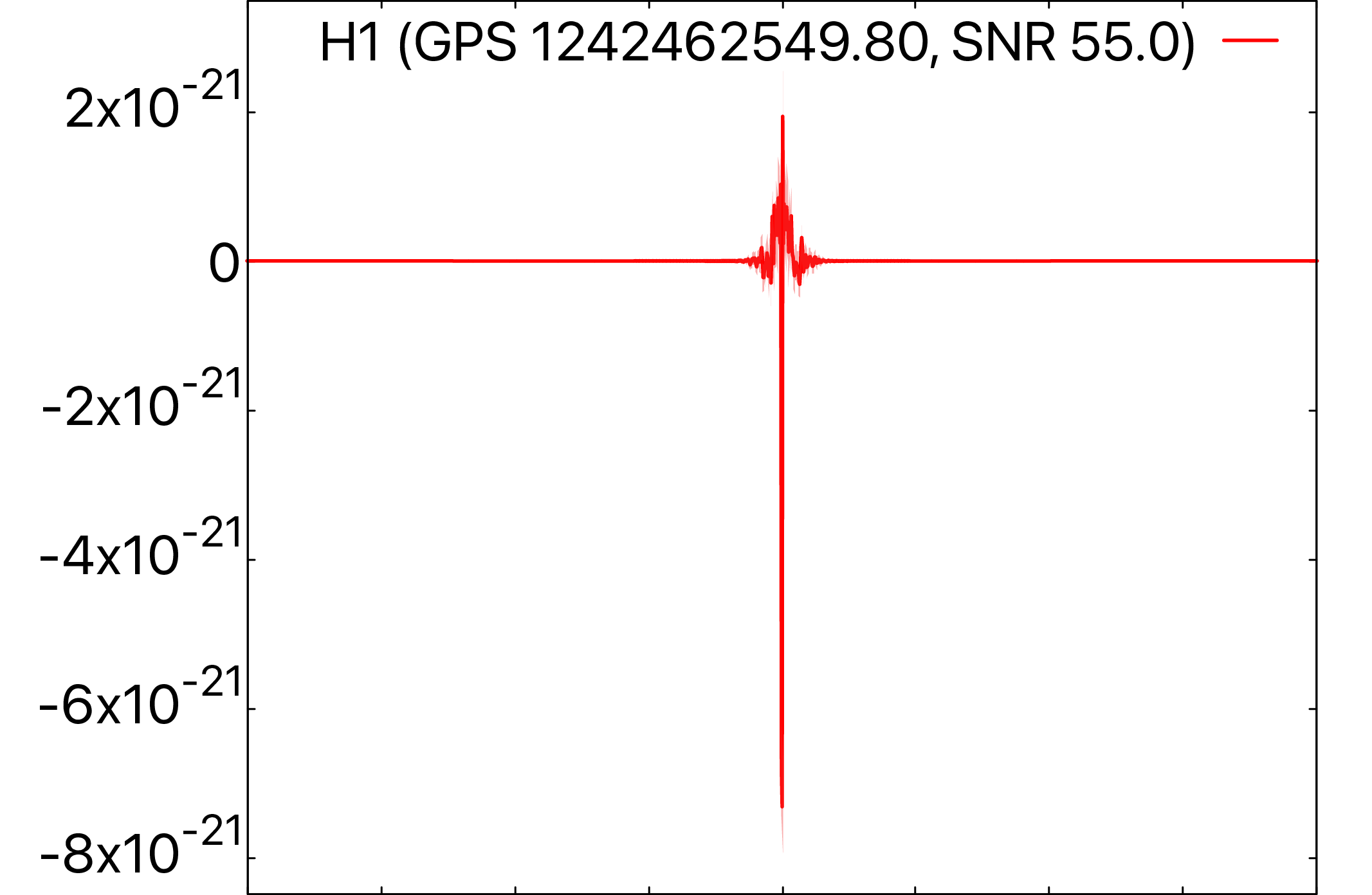}%
    \includegraphics[width=4.4cm]{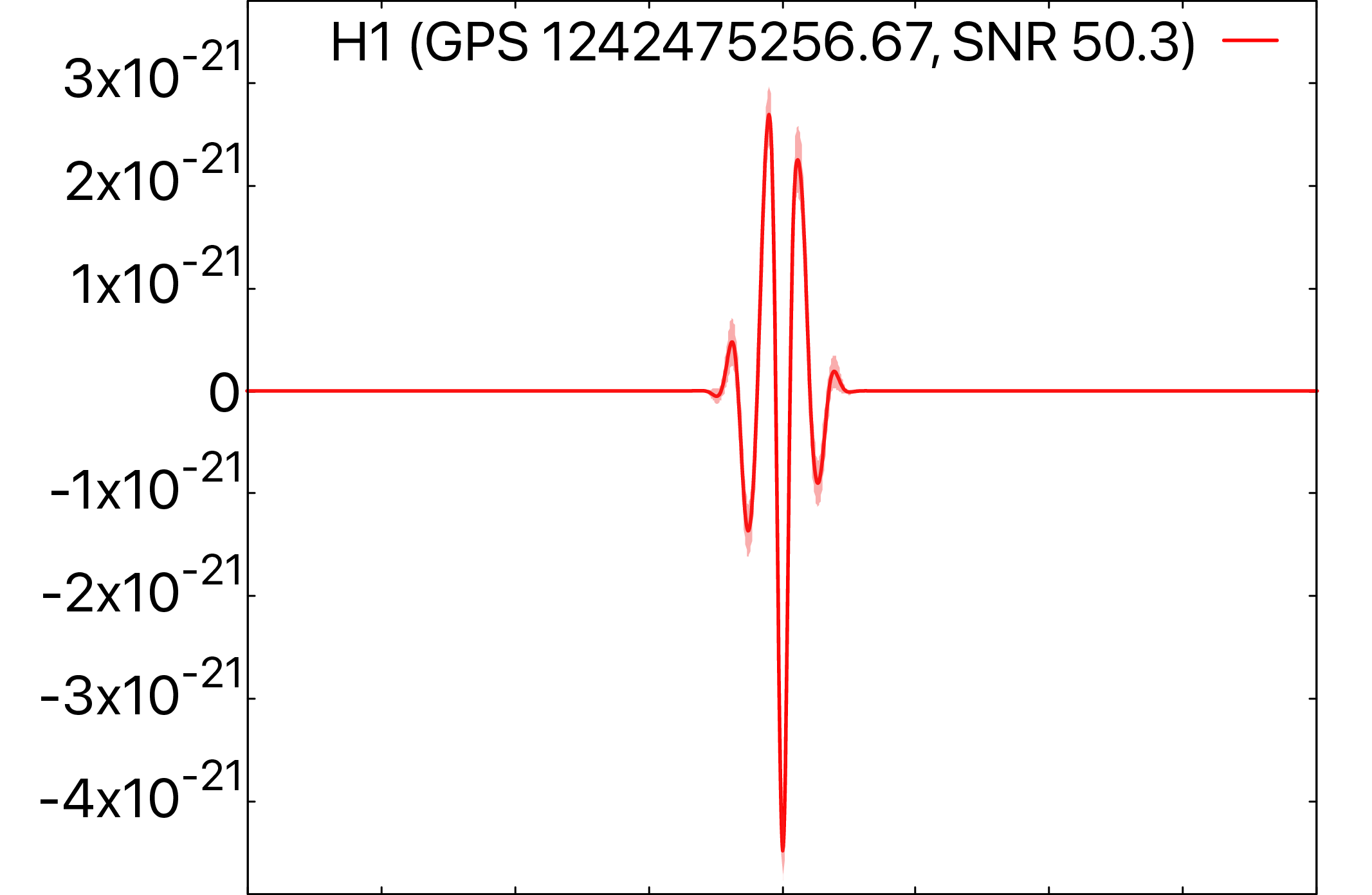}%
    \includegraphics[width=4.4cm]{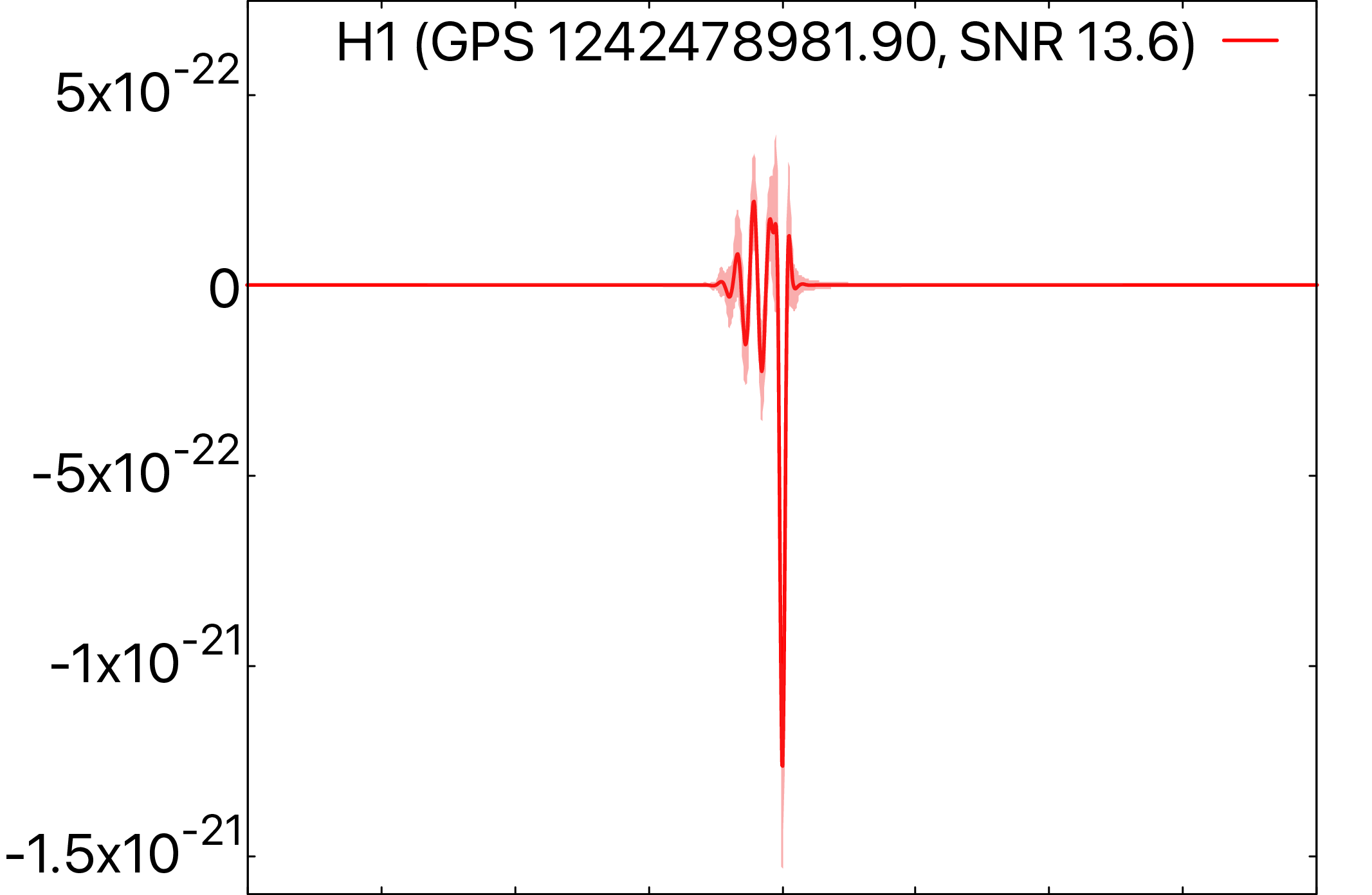}%
    \includegraphics[width=4.4cm]{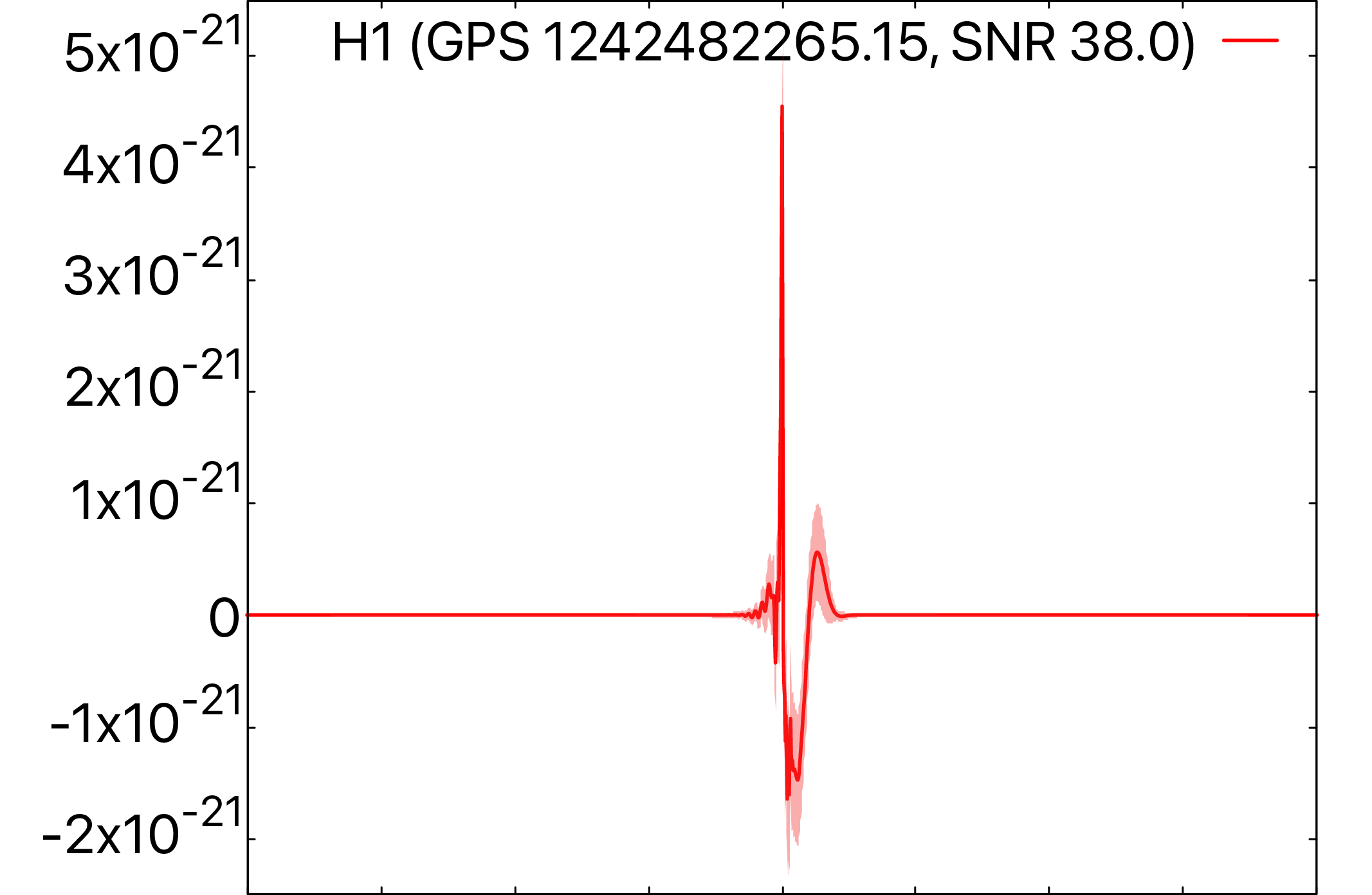}\\
    \includegraphics[width=4.4cm]{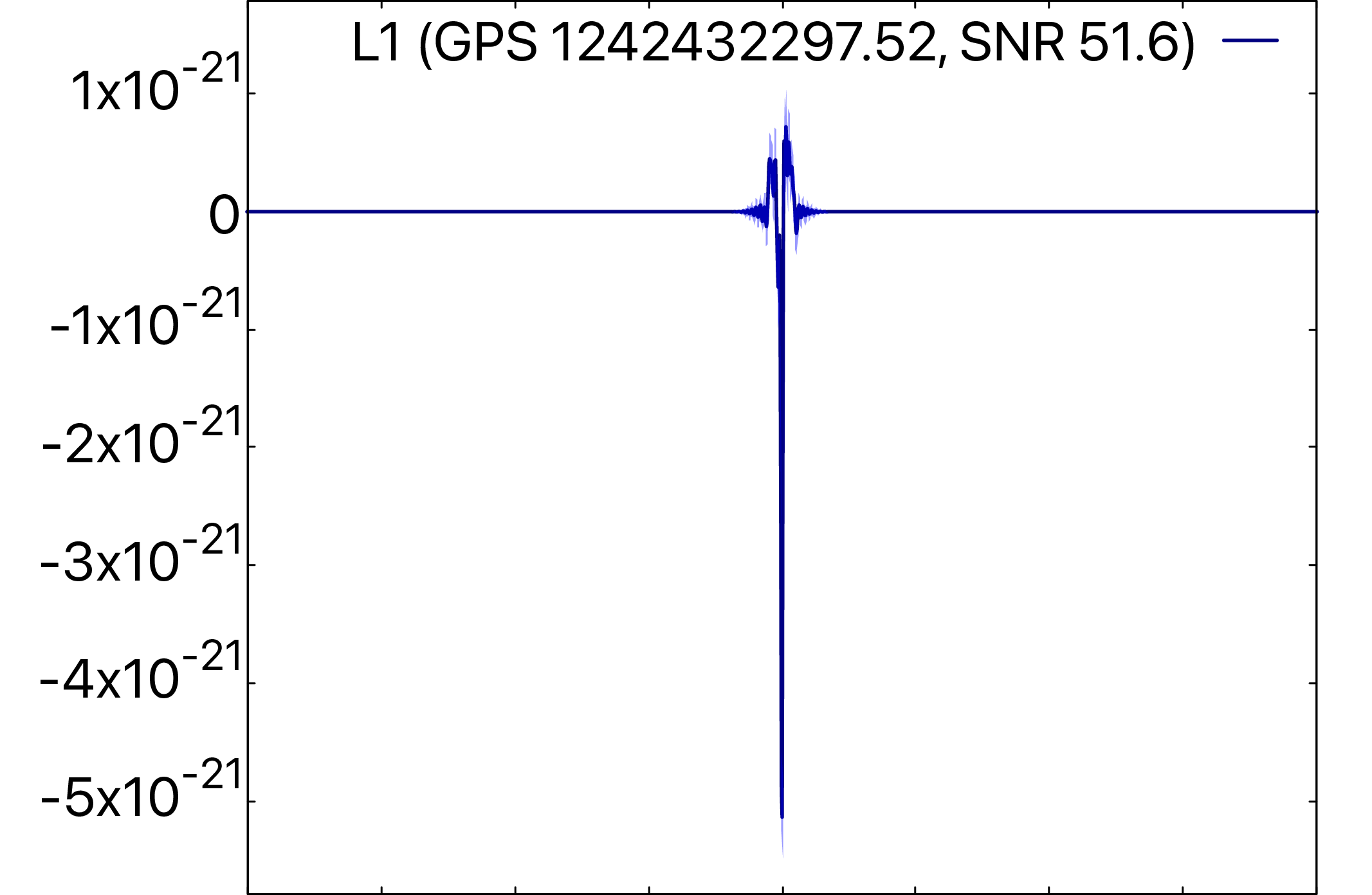}%
    \includegraphics[width=4.4cm]{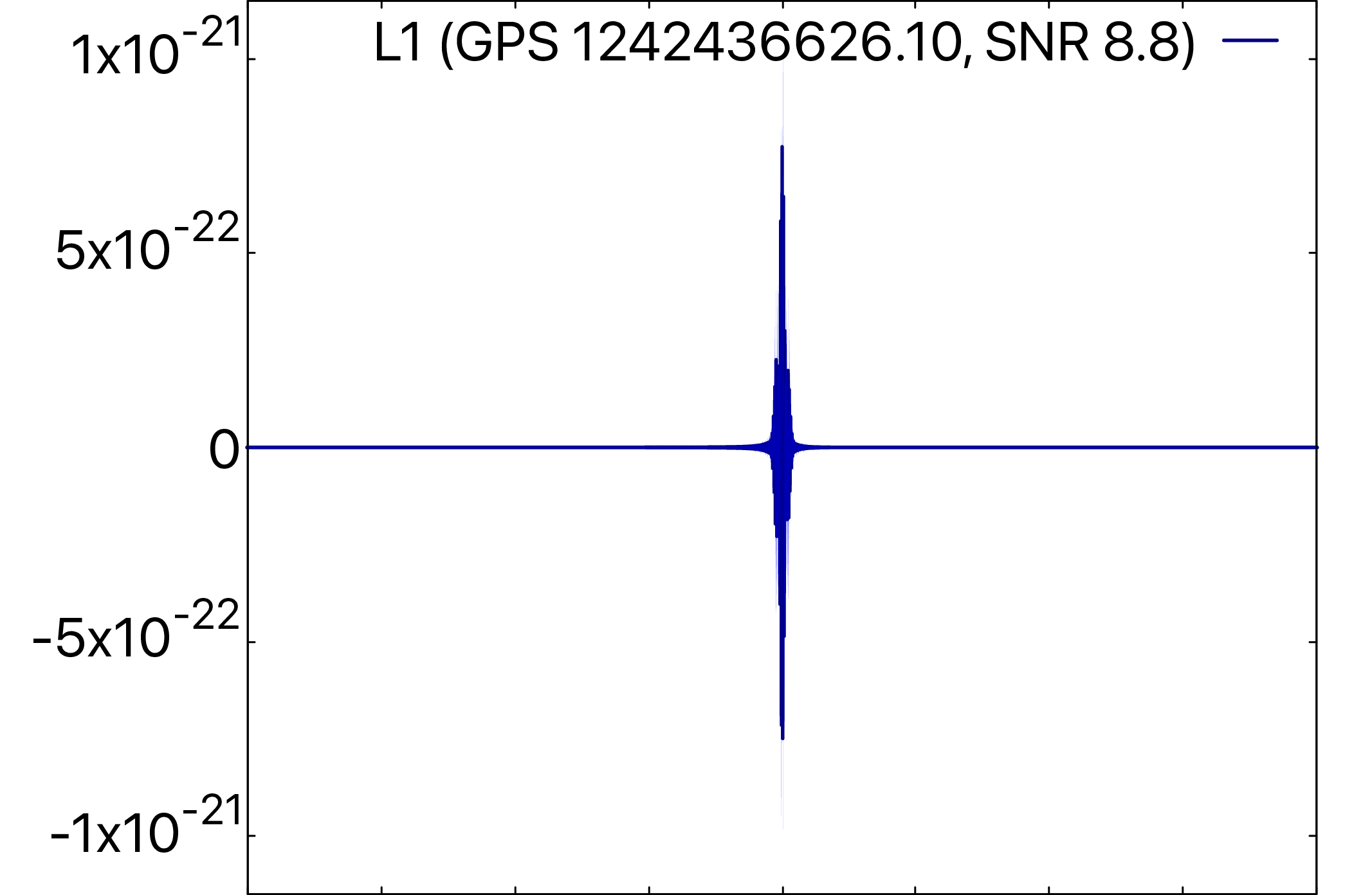}%
    \includegraphics[width=4.4cm]{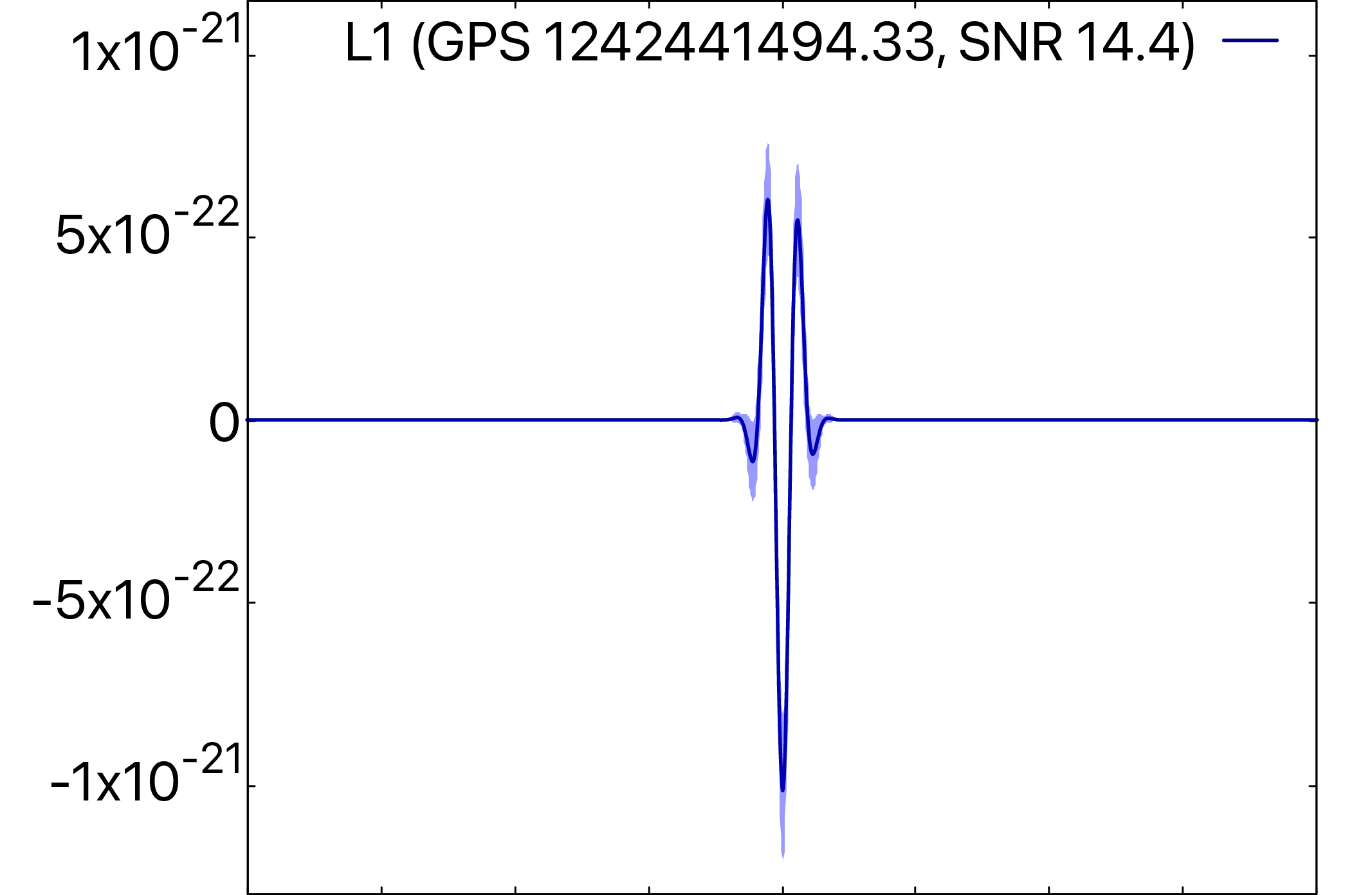}%
    \includegraphics[width=4.4cm]{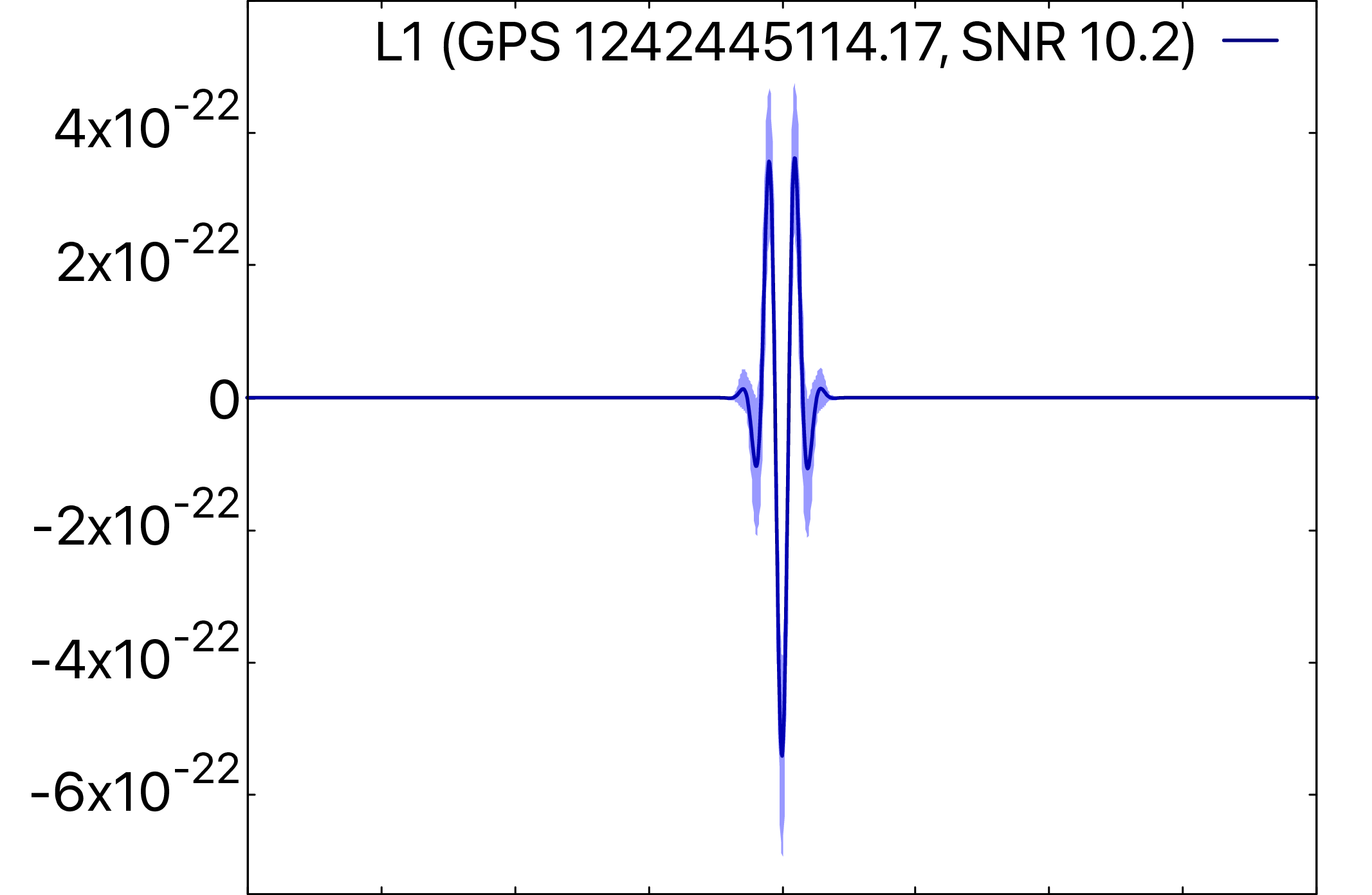}\\
    \includegraphics[width=4.4cm]{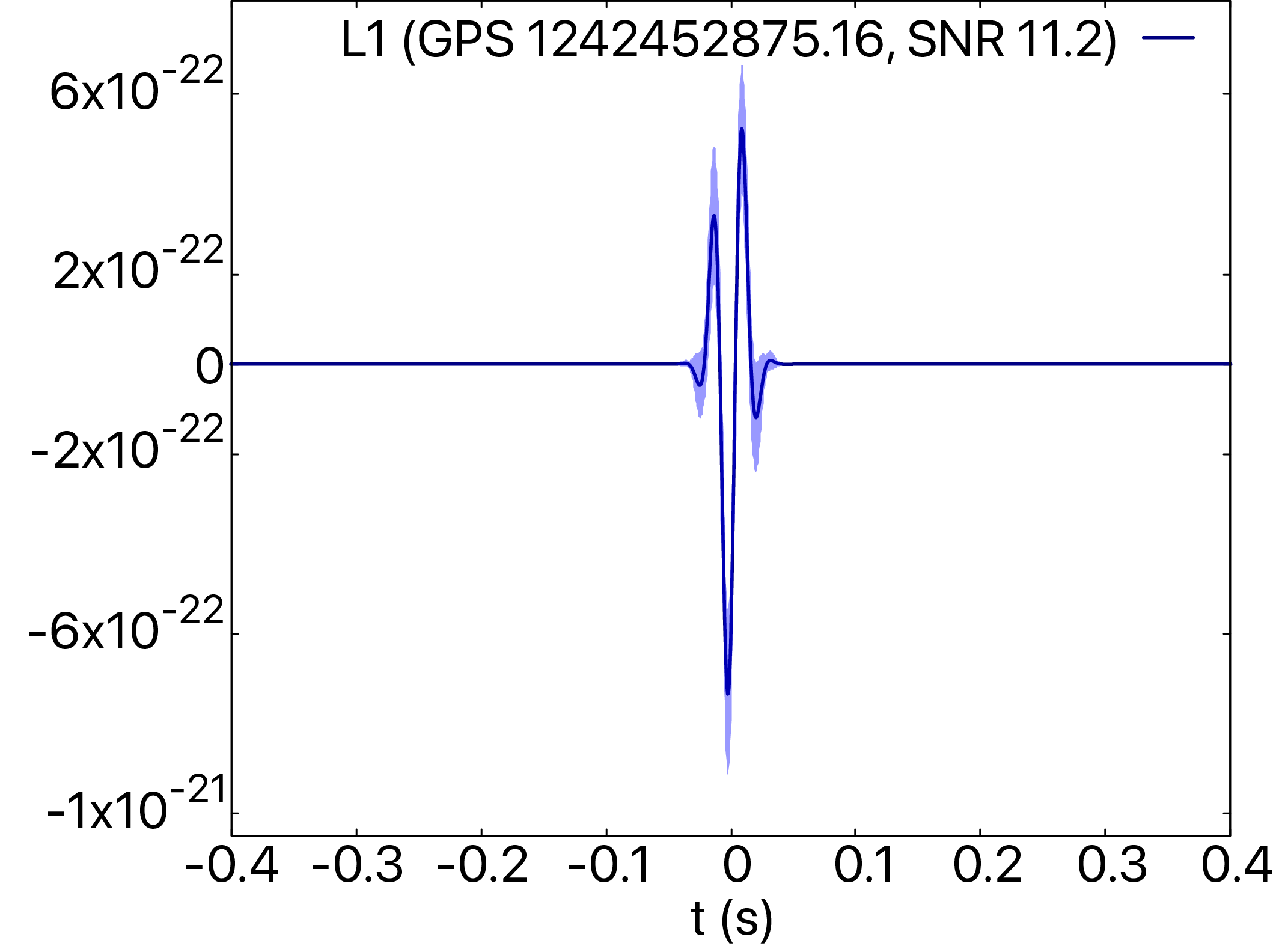}%
    \includegraphics[width=4.4cm]{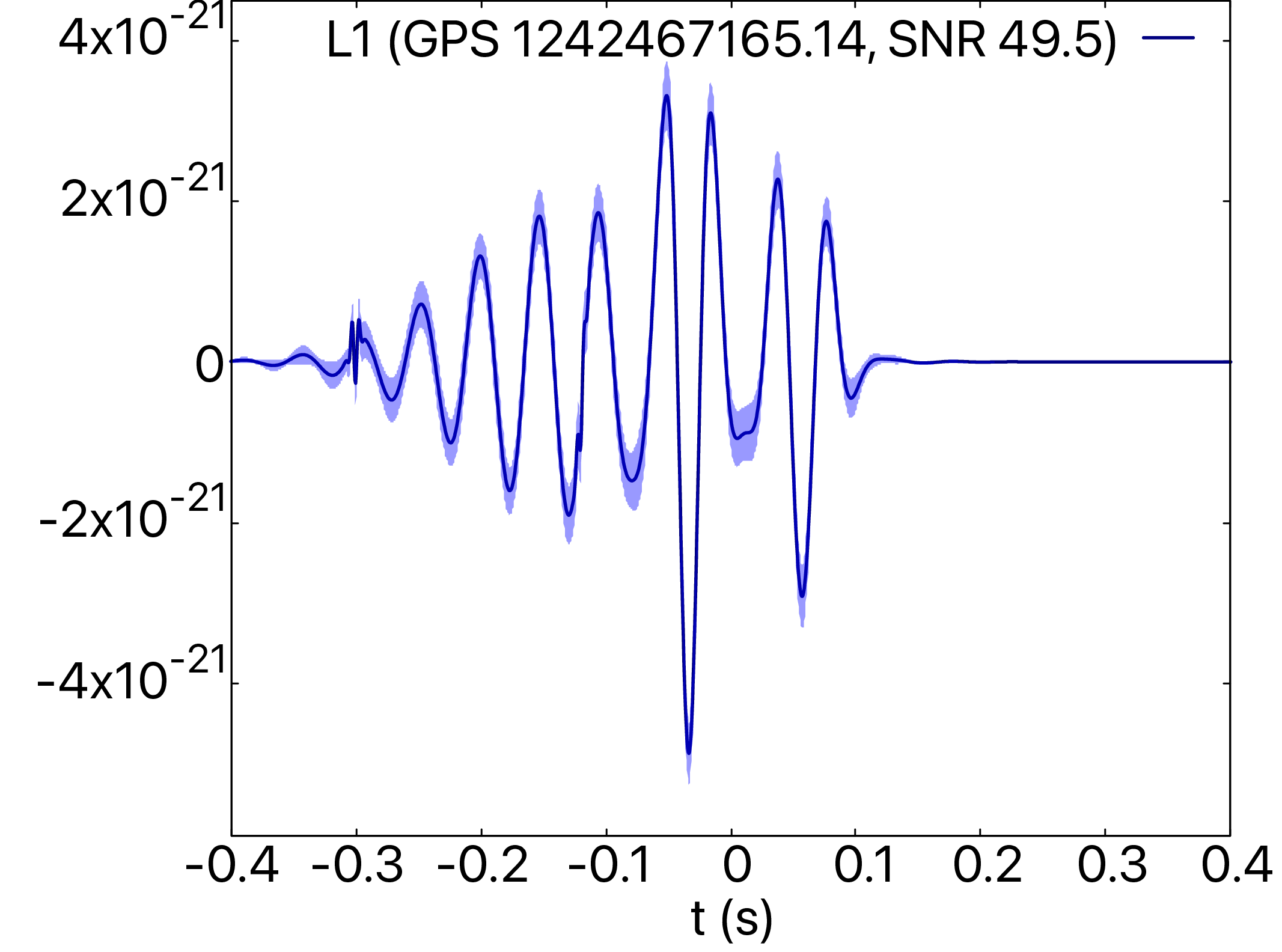}%
    \includegraphics[width=4.4cm]{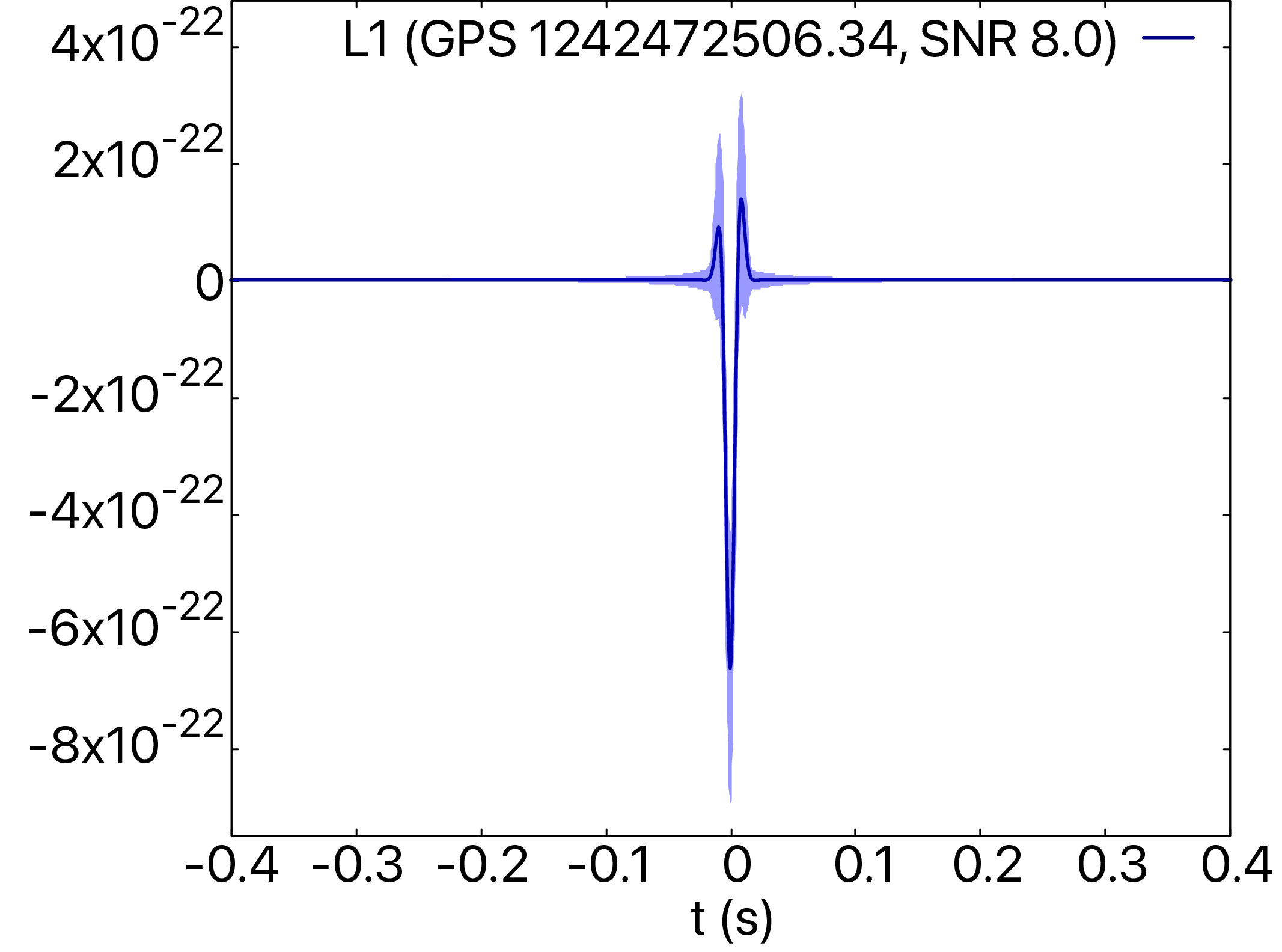}%
    \includegraphics[width=4.4cm]{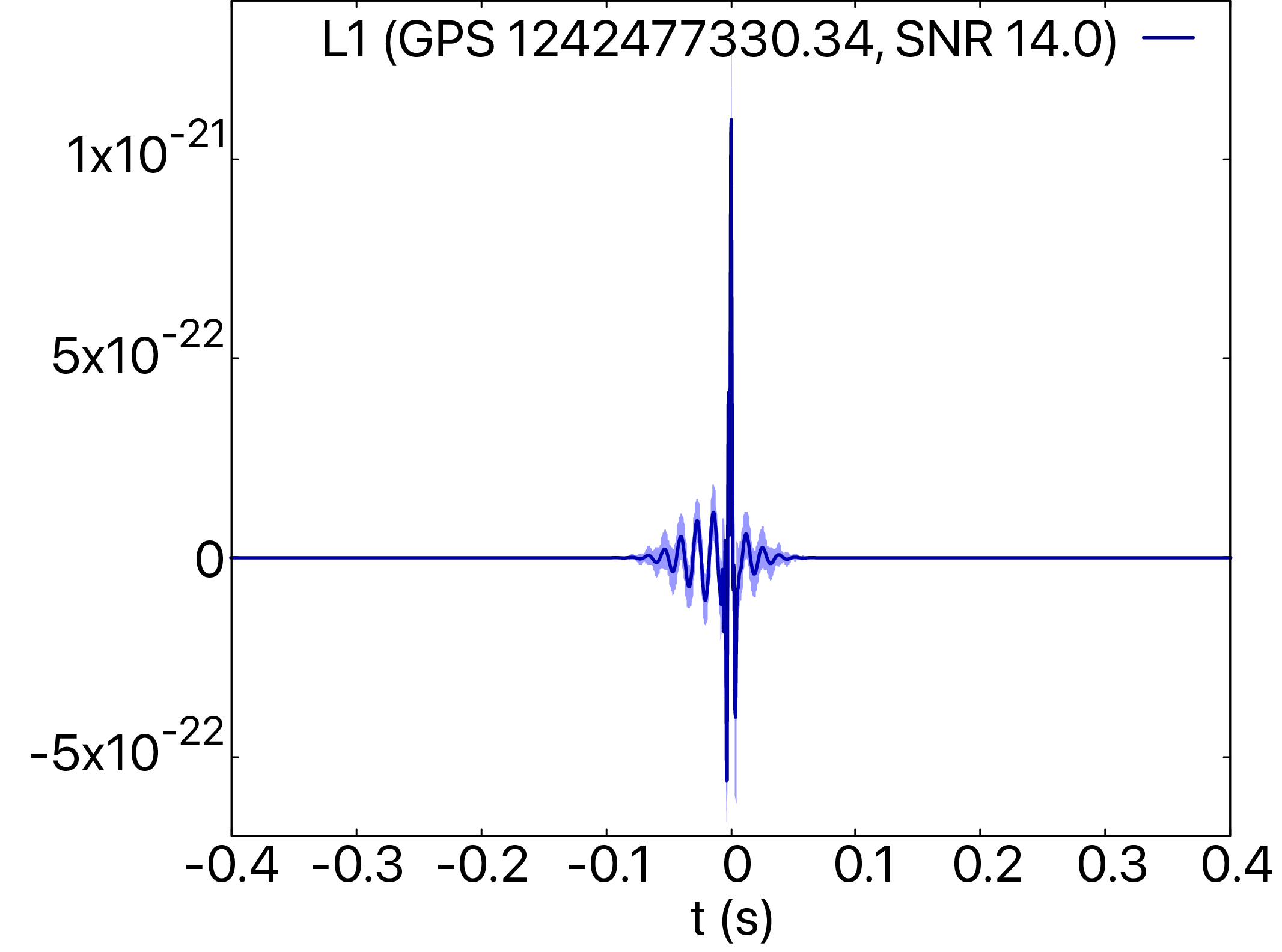}\\
    \caption{Gallery of a few glitches reconstructed by MaxWave in the H1 (red, first and second row) and L1 (blue, third and fourth row) detectors with various SNRs, morphologies, and GPS times for the 14 hour LIGO data run on May 21, 2019 \cite{GWTC-3}.}
    \label{fig:frames_glitch_gallery}
\end{figure*}

We plot histograms of the reconstructed SNR $\in [5,100]$ glitches in the 14 hour H1 and L1 data in Fig. \ref{fig:frame_hist} and present a gallery of H1 (red, first and second row) and L1 (blue, third and fourth row) glitch reconstructions with various SNRs and morphologies at a few different GPS times across the 14 hours in Fig. \ref{fig:frames_glitch_gallery}.

\subsection{Special Case: Glitches overlapping signals}\label{subsec:glitches_overlapping_signals}

\begin{figure}
    \centering
    \includegraphics[width=\linewidth]{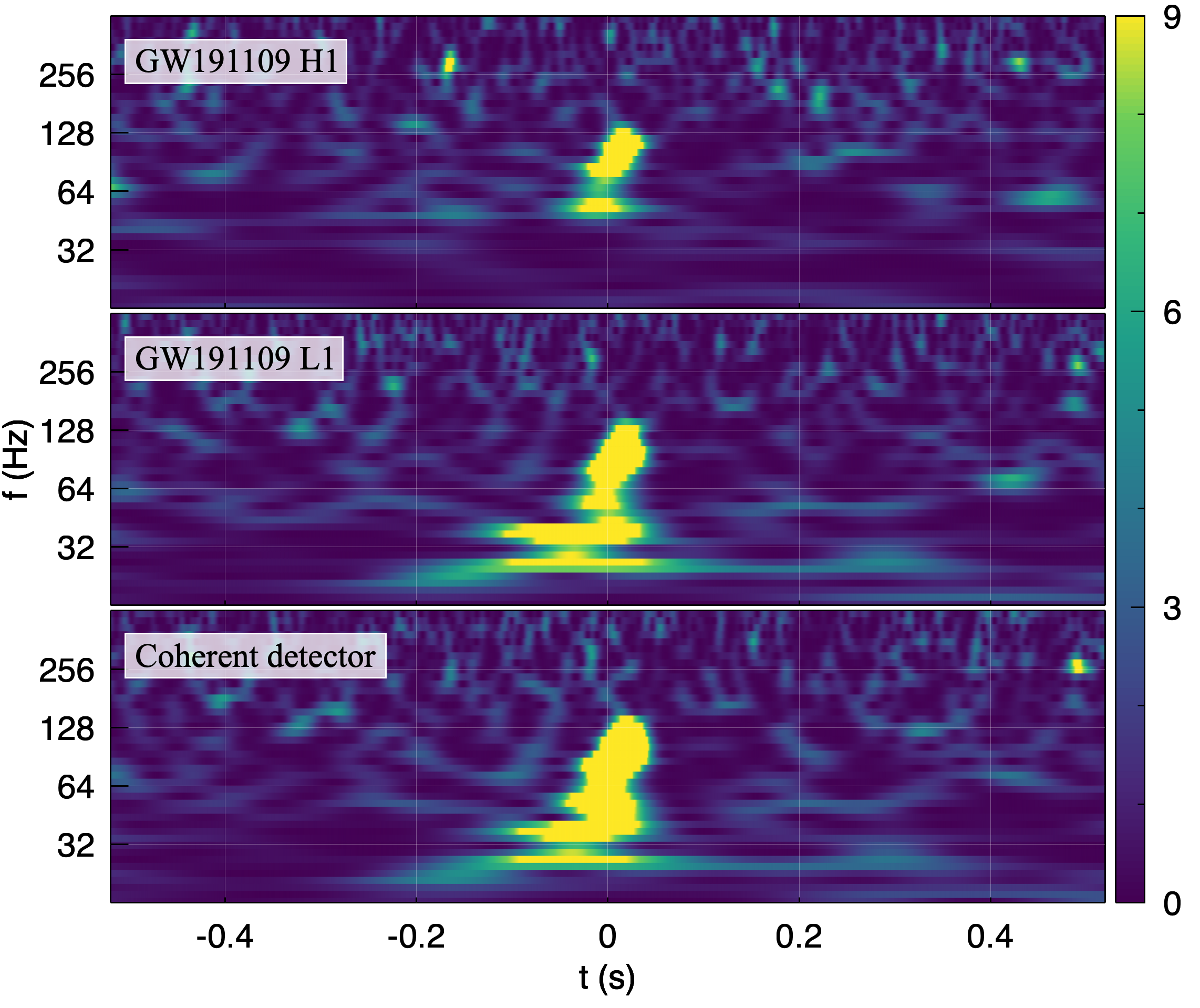}
    \caption{Qscans of the GW191109 event \cite{GWTC-3, GWOSC_O1_O3} at $Q =8$ in the individual LIGO detectors H1 (top), L1 (middle), and in our synthetic coherent detector (bottom). The L1 detector contains a scattered light glitch overlapping with the signal in the low frequency range \cite{Udall_2024}. The coherent detector, by construction, amplifies the signal, reduces incoherent glitch power, and preserves Gaussian noise statistics.}
    \label{fig:GW191109 Qscan}
\end{figure}

We further test our model on gravitational wave events GW191109 and GW200129 in the LIGO O3 data \cite{GWTC-3, GWOSC_O1_O3} where overlapping or nearby glitches affect source parameter inference for the signals. The GW191109 candidate is a high mass binary whose inferred anti-aligned spin \cite{GWTC-3} favors a dynamical formation channel \cite{Zhang_2023} and the GW200129 candidate shows evidence for spin-induced orbital precession \cite{Hannam_2022, Varma_2022}. However, both these inferences strongly depend on how the overlapping or nearby glitches are modeled. For GW191109, when the overlapping glitch in the L1 detector is modeled as slow scattered light, the inference favors anti-aligned spins, while a more flexible glitch model based on sine-Gaussian wavelets yields a bimodal solution with both aligned and anti-aligned spins \cite{Udall_2024}. For GW200129, an instrument glitch in L1 coincides with the frequencies constraining precession \cite{Davis_2022} and the precession evidence is localized to the 20 to 50 Hz L1 data \cite{Payne_2022} and depends on the applied glitch mitigation method \cite{Payne_2022, Macas_2024}. 

\begin{figure}
    \centering
    \includegraphics[width=\linewidth]{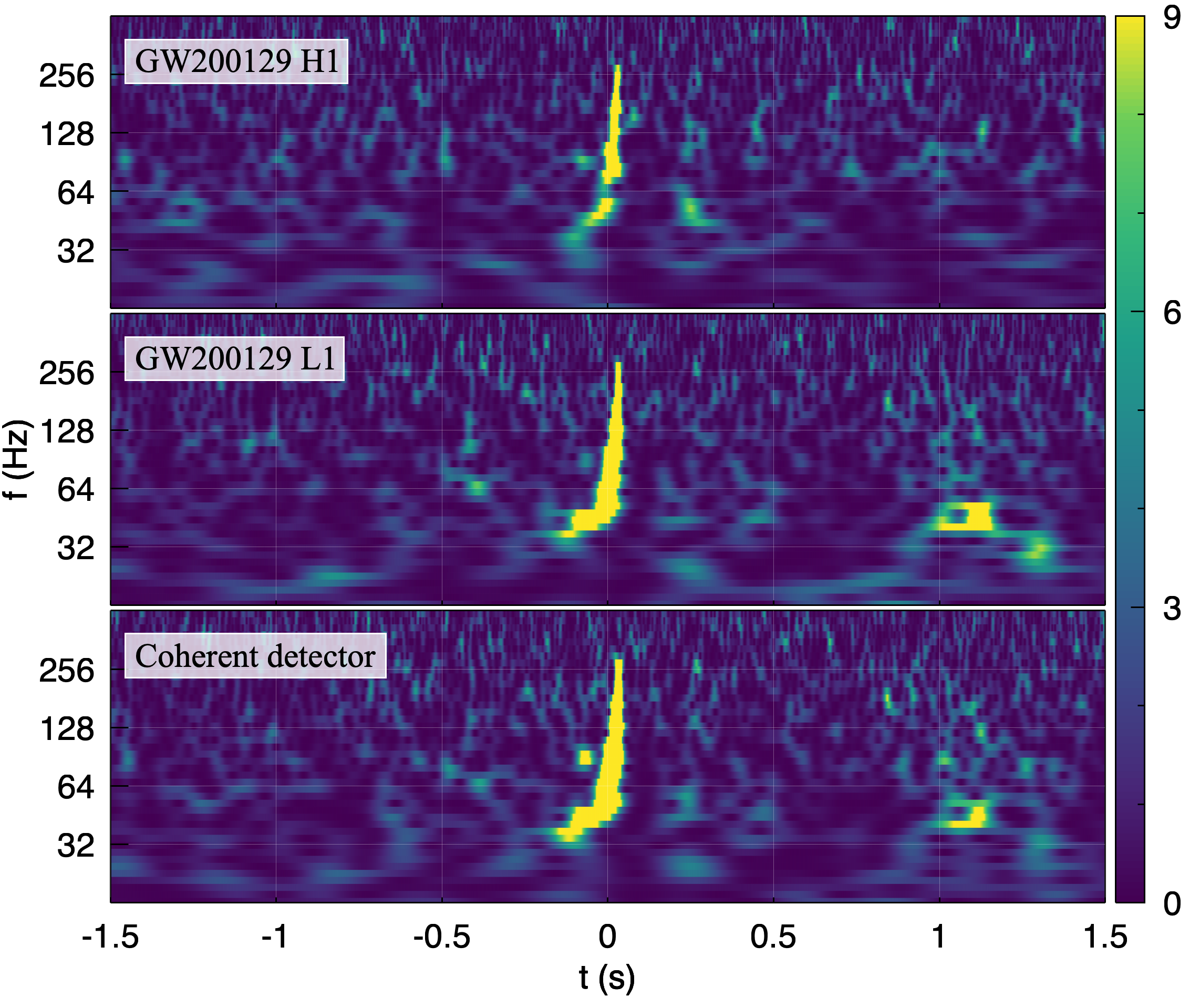}
    \caption{Qscans of the GW200129 event \cite{GWTC-3, GWOSC_O1_O3} at $Q =8$ in the individual LIGO detectors H1 (top), L1 (middle), and in our synthetic coherent detector (bottom). The L1 detector contains an instrument glitch below 70 Hz that coincides with the frequencies constraining spin-induced orbital precession \cite{Davis_2022, Payne_2022, Macas_2024}. The coherent detector visibly reduces the glitch power while amplifying the signal.}
    \label{fig:GW200129 Qscan}
\end{figure}

We plot the Qscans of the GW191109 and GW200129 events (Figs. \ref{fig:GW191109 Qscan} and \ref{fig:GW200129 Qscan} respectively) in the individual LIGO detectors H1 (top), L1 (middle) and in our synthetic coherent detector (bottom). In the GW200129 Qscans, we can visually see that the coherent detector reduces the glitch power while amplifying the signal. 

\begin{figure*}
    \centering
    \includegraphics[width=\linewidth]{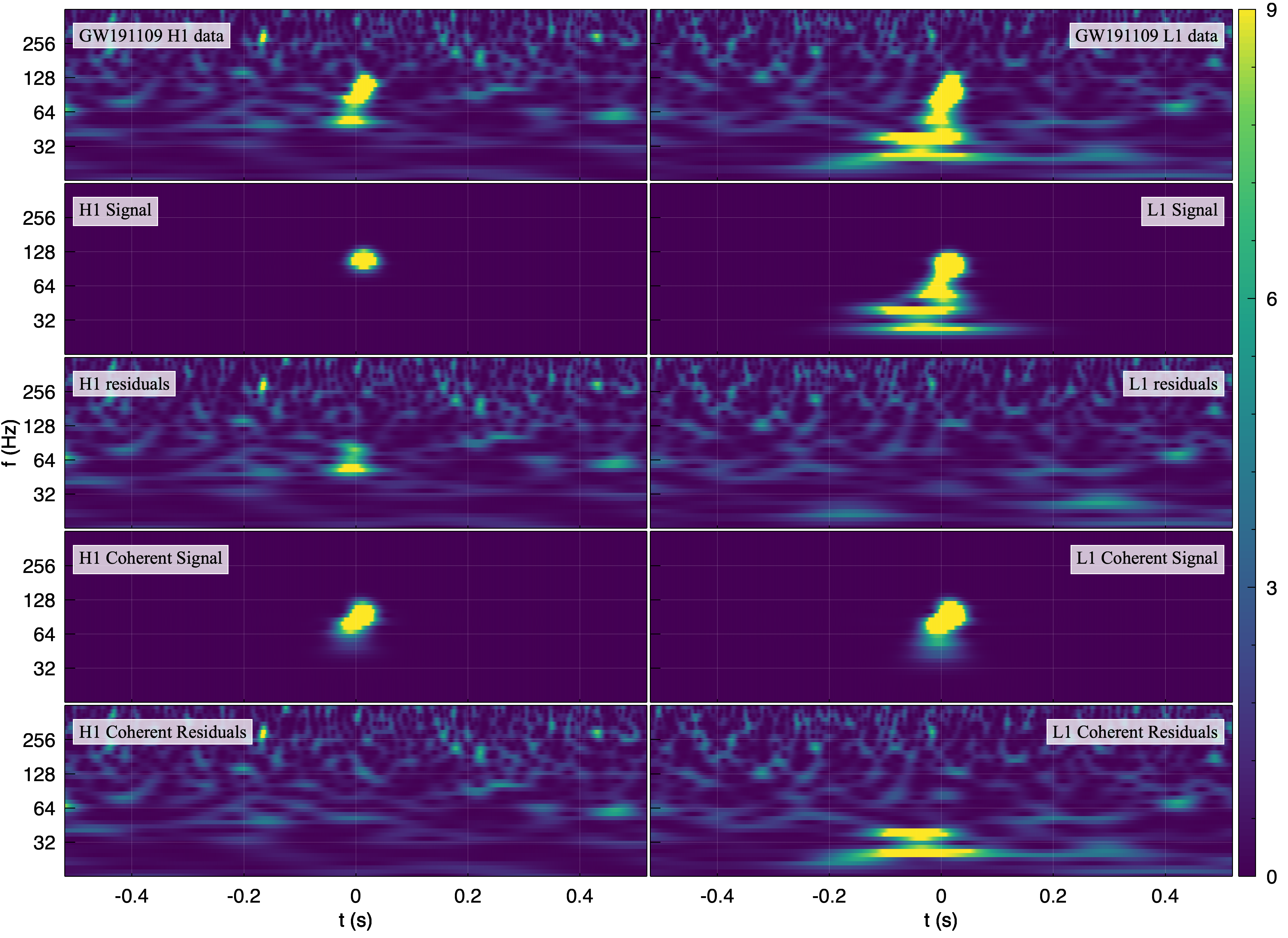}
    \caption{Qscans of the GW191109 event \cite{GWTC-3, GWOSC_O1_O3} at $Q=8$ (top row) in the LIGO H1 (left column) and L1 (right column) detectors, along with the respective individual detector reconstructions (row 2) and residuals (row 3), and shifted coherent reconstructions (row 4) and residuals (row 5). The coherent residuals (row 5/ bottom row) remove the signal power much more effectively as compared to the individual detector residuals (row 3), while retaining the scattered light glitch \cite{Udall_2024} power in the L1 coherent residual (bottom right).}
    \label{fig:GW191109 residuals 2}
\end{figure*}

\begin{figure*}
    \centering
    \includegraphics[width=\linewidth]{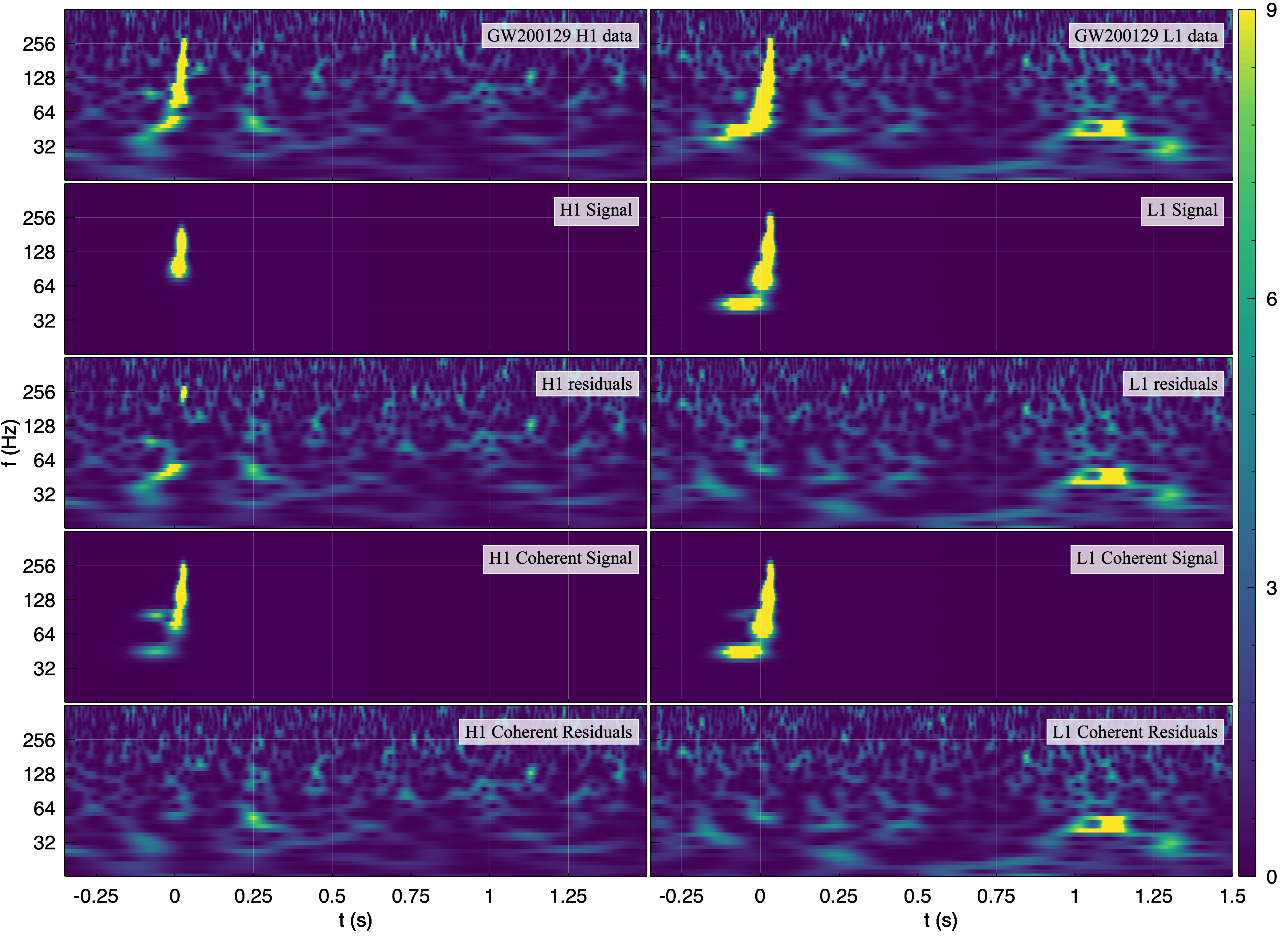}
    \caption{Qscans of the GW200129 event \cite{GWTC-3, GWOSC_O1_O3} at $Q=8$ (row 1) in the LIGO H1 (left column) and L1 (right column) detectors, along with the respective individual detector reconstructions (row 2) and residuals (row 3), and shifted coherent reconstructions (row 4) and residuals (row 5). The coherent residuals (row 5/ bottom row) remove the signal power much more effectively as compared to the individual detector reconstruction residuals (row 3) and retain the instrument glitch \cite{Davis_2022} power in the L1 coherent residual (bottom right).}
    \label{fig:GW200129 residuals 2}
\end{figure*}

We compare the individual and coherent reconstructions (rows 2 and 4) and their corresponding residuals (rows 3 and 5) for the H1 (left column) and L1 (right column) detectors for both the events in Figs. \ref{fig:GW191109 residuals 2} and \ref{fig:GW200129 residuals 2} respectively, where the H1 and L1 data Qscans are also re-plotted for better visualization (row 1). In both cases, the shifted coherent reconstructions (row 4) successfully isolate the glitches and model the signals better than the corresponding individual reconstructions (row 2). By combining multi-detector data, our synthetic detector captures the coherent signal power while the overlapping glitch, which is incoherent across the network, remains in the residual. For GW191109 (Fig. \ref{fig:GW191109 residuals 2}), the coherent reconstruction residuals show no remnant signal power for H1 and isolate the scattered light glitch for L1 (row 5). By contrast, the individual H1 reconstruction leaves signal power in its residual, and the individual L1 reconstruction absorbs the glitch into the recovered signal, leaving a clean residual but contaminating the reconstructed waveform (rows 2 and 3). For GW200129 (Fig. \ref{fig:GW200129 residuals 2}), the coherent residuals show no remnant signal power for H1 or L1 and isolate the instrument glitch in L1 (row 5). The individual H1 reconstruction is less clean and leaves signal power in its residual (rows 2 and 3). These events have also been studied by the deep learning model AWaRe \cite{AWaRe1, AWaRe2}, which reconstructs them using an attention-boosted neural network without explicitly training on glitches, but leaves some signal in the L1 residual for GW200129 \cite{AWaRe3_GW191109_GW200129}. The MaxWave signal model offers a complementary approach based on multi-detector coherence, rather than a learned single detector reconstruction, and isolates the glitch without leaving remnant signal power for either event.

Note that our signal model does not reliably flag the overlapping glitch and signal cases as signals or coincident events, as such cases are beyond the scope of the clean coherent residuals test. Whether an overlapping glitch and signal case passes or fails the clean coherent residuals test depends on the power and the time-frequency spread of the glitch left behind in the residual, which determines whether $\tilde{h}_\text{resid}^n(f)$ crosses the NW $> 1$ threshold. For these two events, GW191109 fails the test and is flagged as a coincident event, while GW200129 passes and is flagged as a signal. Both flags are non-removals, so neither event is removed during cleaning. The strength of the model in this regime is the reconstruction, which isolates the signal while mitigating the overlapping glitch. Regardless of the coincident event or signal non-removal flag, the coherent reconstruction remains informative and provides a good initial glitch-mitigated solution for computationally expensive algorithms such as \textit{BayesWave} \cite{Cornish_2021}, which can then perform comprehensive CBC+Glitch analysis \cite{Sophie-2022}.

\section{CONCLUSION}\label{sec:Conclusion}

We extend MaxWave, a fast maximum likelihood wavelet reconstruction algorithm, to perform coherent multi-detector signal reconstruction and glitch rejection. By aligning individual detector data using z-statistic time and phase shifts, and amplitude scalings, we construct a synthetic coherent detector with preserved Gaussian noise statistics and extract signal information inaccessible to single-detector analyses. Applied to GW150914, the coherent detector amplifies signal power and produces cleaner residuals in both detectors. For shifted coherent detector recoveries of IMRPhenomD binary black hole signals injected in real LIGO noise for H1 and L1 detectors, we observe a $6.3_{-1.7}^{+3.3} \%$ improvement in the match with the injected template at network SNR 10 and $3.2_{-0.7}^{+1.2} \%$ improvement at network SNR 20 as compared to the single-detector recoveries. The coherent signal model converges towards \textit{BayesWave} two detector reconstruction accuracies at higher SNRs and individual binary masses, indicating that the coherent reconstruction is a good initial solution for the \textit{BayesWave} signal model's RJMCMC sampler. The z-statistic light travel time test alone rejects 74.4\% of artificially simultaneous glitches across 10 glitch categories and the coherent residuals test raises the rejection rate to 94.0\%, with cross-type coincidences rejected at 95.9\%. 

When run on 14 hours of real LIGO O3 data \cite{GWTC-3}, MaxWave signal correctly identifies the gravitational wave events GW190521\_030229 and GW190521\_074359 with SNRs of 14.8 and 25.1, consistent with the GWOSC catalog values \cite{GWOSC_O1_O3}, and achieves a time-slide test FAR of $\leq 1.99 \times10^{-5}$ Hz, which corresponds to less than two false alarms in a day. The signal model also maintains real-time performance for networks of up to five detectors, with 50 perturbative refinements of the shifted coherent reconstruction, by running in less than 3.4 s for $T_\text{obs} = 4$ s and $N_\tau = 8$ layers. Our model's real-time coherent signal reconstruction, promising glitch rejection, and low FAR motivate its future development into a complete, stand-alone burst search pipeline.

We further apply the model to the overlapping glitch and signal events GW191109 and GW200129, where the coherent residuals remove the signal power far more effectively than the individual detector residuals, while retaining the glitch power. Although the coherent residuals test is not designed for such overlapping cases, the glitch-mitigated coherent reconstruction can provide a good initial solution for computationally expensive algorithms like the \textit{BayesWave} signal \cite{Cornish_2021} and CBC+Glitch model \cite{Sophie-2022} and speed the convergence times of their comprehensive analysis.

\section{FUTURE DIRECTIONS AND SCOPE}\label{sec:Future Directions and Scope}
We can apply and extend our algorithm for the rapid, low-latency reconstruction of transient signals in multiple directions. This includes safely denoising gravitational wave data without removing transient signals.

MaxWave signal's coherent multi-detector reconstruction can also provide an approximate maximum likelihood initialization and accelerate convergence times for the \textit{BayesWave} signal model \cite{Cornish_2021}, which currently draws all wavelet and extrinsic parameters from broad flat prior distributions. Unlike the \textit{BayesWave} noise model \cite{Cornish_2015, MaxWave_Glitch}, no fast initialization exists for the signal model, as it requires a geocentric waveform coherently projected across the detector network with jointly constrained sky location, polarization, and ellipticity. Our model directly provides this geocentric starting point, seeding the RJMCMC chains near a good solution in an otherwise hard-to-sample, multimodal parameter space. The central time, central frequency, and time extent of the wavelets identified by our coherent reconstruction can be used as a proposal to the \textit{BayesWave} signal model \cite{Cornish_2021}, encouraging it to put down a wavelet where there is coherent power. To make our proposal legal, it must have a finite proposal density. We can accomplish this by constructing the prior on the amplitude and phase of the proposed wavelets to be a Gaussian blob around the approximate maximum likelihood solution.

Our model currently assumes elliptical polarization, which is suboptimal for unpolarized signals such as white noise bursts and for highly precessing CBCs with time-varying polarization content. We intend on relaxing our assumption to incorporate generic polarizations. Similar to the \textit{BayesWave} generic polarization model \cite{Cornish_2021}, we can construct the polarization modes $h_+$ and $h_\times$ with independent sets of wavelets that differ in $\mathcal{A}^n, \phi_0^n$ but share the $t_0^n, f_0^n,$ and $\tau^n$ parameters.

MaxWave signal diversifies burst search and reconstruction algorithms like cWB \cite{cwb1, cwb2}, X-Pipeline \cite{Xpipeline}, DeepExtractor \cite{DeepExtractor}, and AWaRe \cite{AWaRe1, AWaRe2} by offering a conceptually distinct framework, strengthening detection confidence and improving sensitivity to diverse signal morphologies. It also complements astrophysical model-based and computationally expensive algorithms like \textit{BayesWave} \cite{Cornish_2021}, CBC+Glitch \cite{Sophie-2022} and QuickCBC \cite{Neil_QuickCBC}, and can provide real-time, model-independent signal reconstruction and glitch rejection. By further quantifying the MaxWave signal's detection statistics across the signal parameter space, and comparing its performance to that of the existing burst search algorithms, our model can be extended to a complete, stand-alone low-latency burst search pipeline capable of processing entire LIGO-Virgo-KAGRA observation runs \cite{LVK} in real time.

\section*{ACKNOWLEDGMENTS}
This work was supported by NSF Award Numbers PHY-2207970 and PHY-2513363. The authors thank Shobhit Ranjan for help with setting up the \textit{BayesWave} runs, and Dr. Sophie Bini and Dr. Chayan Chatterjee for their insightful comments. The authors are grateful for computational resources provided by the LIGO Laboratory and supported by National Science Foundation Grants No. PHY-0757058 and No. PHY-0823459. The data used in our study is based upon work supported by NSF's LIGO Laboratory which is a major facility fully funded by the National Science Foundation.

\bibliography{bibliography_short} 

\end{document}